\documentclass[pdflatex,sn-mathphys-num]{sn-jnl}% Math and Physical Sciences Numbered Reference Style
\usepackage{graphicx}%
\usepackage{multirow}%
\usepackage{amsmath,amssymb,amsfonts}%
\usepackage{amsthm,bm}%
\usepackage{mathrsfs}%
\usepackage[title]{appendix}%
\usepackage{xcolor}%
\usepackage{textcomp}%
\usepackage{manyfoot}%
\usepackage{booktabs}%
\usepackage{algorithm}%
\usepackage{algorithmicx}%
\usepackage{algpseudocode}%
\usepackage{listings}%

\graphicspath{ {./} }
\usepackage{array}
\usepackage{subfig}
\usepackage{tabularx}

\theoremstyle{thmstyleone}%
\theoremstyle{thmstyletwo}%

\theoremstyle{thmstylethree}%

\begin{document}

\title[Surface Dynamics About Comet 67P/Churyumov-Gerasimenko]{Equilibrium Points and Surface Dynamics About Comet 67P/Churyumov-Gerasimenko}


\author*[1]{\fnm{Leonardo} \sur{Braga}}\email{lb.braga@unesp.br}

\author[1]{\fnm{Andre} \sur{Amarante}}

\author[1]{\fnm{Alessandra} \sur{Ferreira}}

\author[1]{\fnm{Caio} \sur{Gomes}}

\author[1]{\fnm{Luis} \sur{Ceranto}}

%\affil*[1]{\orgdiv{Department of Mathematics}, \orgname{\textbf{São Paulo State University (UNESP), School of Engineering and Sciences, Guaratinguetá}}, \orgaddress{\street{Av. Dr. Ariberto Pereira da Cunha, 333}, \city{Guaratinguetá}, \postcode{12516-410}, \state{SP}, \country{Brazil}}}
\affil[1]{\orgdiv{Department of Mathematics}, \orgname{São Paulo State University (UNESP), School of Engineering and Sciences}, \city{Guaratinguetá}, \state{SP}, \country{Brazil}}


\abstract{
%\textbf{(In this study, we analyzed the surface dynamics of comet Churyumov-Gerasimenko (67P) in detail, utilizing a high-resolution 3-D polyhedral shape model. We use the polyhedral method to compute dynamical surface characteristics such as geometric height, surface tilt, surface slopes, geopotential surface, acceleration surface, escape speed, equilibrium points, and zero-velocity curves. In addition, we assume that comet 67P is a dipole-segment model to study the evolution of the characteristic curves around it. Our results show that the gravitational potential dominates the geopotential surface due to the comet's slow rotation. The escape speed has higher intensities in the Hapi region (neck). The surface slopes are analyzed to predict possible particle movement and accumulation regions. Also, we identified five equilibrium points around comet 67P, where equilibrium point E$_2$ is linearly stable. Finally, we found 12 families of symmetric periodic orbits around the comet 67P using the dipole-segment approach. The detailed analysis presented in this work helps understand the complex gravitational environment of comet 67P dynamically, providing essential information for future exploration efforts.)} \\
Small bodies in our Solar System are considered remnants of their early formation. Studying their physical and dynamic properties can provide insights into their evolution, stability, and origin. ESA's Rosetta mission successfully landed and studied comet Churyumov-Gerasimenko (67P) for approximately two years. In this work, the aim is to analyze the surface and orbital dynamics of comet 67P in detail, using a suitable 3-D polyhedral shape model. We applied the polyhedron method to calculate dynamic surface characteristics, including geometric height, surface tilt, surface slopes, geopotential surface, acceleration surface, escape speed, equilibrium points, and zero-velocity curves. The results show that the gravitational potential is predominant on the comet's surface due to its slow rotation. The escape speed has the maximum value in the Hapi region (the comet's neck). The surface slopes were analyzed to predict possible regions of particle motion and accumulation. The results show that most regions of the comet's surface have low slopes. Furthermore, we analyzed the slopes under the effects of Third-Body gravitational and Solar Radiation Pressure perturbations. Our results showed that the effects of Third-Body perturbations do not significantly affect the global behavior of slopes.
Meanwhile, the Solar Radiation Pressure does not significantly affect particles across the surface of comet 67P with sizes $>\sim10^{-3}$\,cm at apocenter and $>\sim10^{-1}$\,cm at pericenter. We also identified four equilibrium points around comet 67P and one equilibrium point inside the body, where points E$_2$ and E$_5$ are linearly stable. In addition, we approximated the shape of comet 67P using the simplified Dipole Segment Model to study its dynamics, employing parameters derived from its 3-D polyhedral shape model. We found 12 families of planar symmetric periodic orbits around the body.
}




\maketitle

\section{Introduction}
\label{sec:1}
The Rosetta mission, led by the European Space Agency (ESA), aimed to study the comet Churyumov-Gerasimenko (67P) through remote sensing and in situ measurements. The spacecraft entered orbit around the comet and collected data about its environment and structure. The mission tracked the evolution of the comet along its trajectory, studying changes in its activity before, during, and after its closest approach to the Sun, and concluded with a controlled landing on the comet's surface\footnote{\url{https://www.esa.int/Enabling_Support/Operations/Rosetta}}.
Comets are considered time capsules containing material from our Solar System's early stages. These small bodies, where temperatures are low, are understood to be the least processed bodies in the Solar System, offering a unique insight into the planetary formation.

%\textbf{Preusker et al. \cite{preusker2015} presents the construction of a detailed 3-D model of the comet's nucleus, based on stereo photographic techniques applied to images acquired by the Rosetta spacecraft's Optical, Spectroscopic, and Infrared Remote Imaging System (OSIRIS), where more than 200 images from the Narrow Angle Camera (NAC) were processed, with resolutions ranging from 0.9 to 2.4 meters per pixel, which were acquired between August and September 2014, during the comet's approach and initial mapping phases. The generated model has approximately 16 million faces. With this, 

Preusker et al. \cite{preusker2015} determined the total volume of the nucleus as approximately 18.7 km$^{3}$ by extrapolation from the volume of the comet's northern hemisphere. Pätzold et al. \cite{paztold2016} presented the determination of the mass, average density, and internal structure of the nucleus of comet 67P, based on the analysis of its gravitational field derived from the probe's radio science data. %From perturbations in the spacecraft's velocity during the approach maneuvers, it was possible to determine its total mass of 9.982 $\pm$ 3$\times$10$^{9}$ kg, furthermore combined the mass obtained with the volume obtained by \cite{preusker2015} and obtained an average density of 0.533 g\,cm$^{-3}$.}
Vallat et al. \cite{vallat2017} described in detail the scientific planning process of the mission, from the conditions of the planning of operations, such as thermal constraints, limitations of solar energy generation, gas drag forces, to the orbital conditions resulting from the gravity of its irregular nucleus.

%However, the chemical composition of comets may have changed over time due to the thermal and physical processes they have undergone to date. Thus, measuring the volatile inventory of chemical species in comets and comparing it to other bodies in the Solar System and interstellar matter is essential for understanding primordial materials and formation processes, as well as for assessing the exogenous delivery of water, carbon, and prebiotic molecules to early Earth \cite{leroyetal2015}.
%The Rosetta mission also collected images of asteroids such as (2867) Steins and (21) Lutetia in 2008 and 2010, respectively \cite{Glassmeieretal2007, Rolletal2016}. Other space missions, such as New Horizons (2006), studied the dwarf planet Pluto and the Kuiper Belt Object (KBO)(486958) Arrokoth \cite{stern2007, Sternetal2019}, and JAXA's Hayabusa2 mission \cite{Hirabayashietal2021} collected samples from the asteroid (162173) Ryugu, further expanding the understanding of bodies in the asteroid belt.

Recent studies have examined the connection with surface behaviors and topographic features of celestial bodies like (101955) Benu, (486958) Arrokoth, Ryugu, and (3200) Phaethon, from dynamical characteristics such as geopotential surface, slopes, escape speed, and equilibrium points \cite{scheeresetal2016,Amarante2021,amaranteandwinter2020,Amarante2022,fuetal2024,joetal2025}. 

Scheeres et al. \cite{scheeresetal2016} investigated dynamical features on Bennu's surface and found that particles near the pole could reach velocities that would remain below the asteroid's escape speed.
Amarante \& Winter \cite{amaranteandwinter2020} studied the dynamics of the KBO Arrkoth and identified that the equatorial regions of the two lobes of the contact binary act as geopotential maxima and maintain their surface acceleration between 0.5 and 1 mm s$^{-2}$.
Amarante et al. \cite{Amarante2021} found that large particles could collide with the surface of Bennu at low speeds, preferably in high latitude regions, close to the equator, and also to the north pole of the asteroid. In contrast, areas of latitude would be less populated.
Fu et al. \cite{fuetal2024} identified six external equilibrium points, located along the ridge of Ryugu, and also analyzed their topological stability. They also mapped the families of symmetric periodic orbits associated with these points, identifying 15 families, some of which converge to other equilibrium points. In contrast, others end on the surface of Ryugu, and one of them exhibits a cyclic behavior throughout its evolution.
Jo et al. \cite{joetal2025} analyzed the dynamic environment near the asteroid Phaethon, to support JAXA's DESTINY+ mission, determining several dynamic properties of its surface from its geopotential, such as the escape speed at its surface, which is approximately 1 m\,s$^{-1}$. They also found seven equilibrium points, one of which is linearly stable, and the others are all unstable. 
%For comet 67P, the surface dynamics highly depend on its irregular shape, and several studies have revealed important insights into the dynamics of its gas ejection.

Hasselmann et al. \cite{hasselmann2019} analyzed the Khonsu region, in the southern hemisphere of the comet 67P, also through images provided by the OSIRIS instrument. This region showed intense activity during the comet's pericenter passage. Moreno et al. \cite{morenoetal2022} examined the rotational dynamics of irregularly shaped cometary particles, considering the combined effects of gas drag and radiative forces and torques. Their model applied to comet 67P revealed that radiative torques were negligible compared to gas-driven torques for particles larger than 10\,$\mu$m. Frattin et al. \cite{Frattietal2021} focused on the motion of dust particles in the inner coma of 67P, analyzing OSIRIS images to study the trajectories, rotation, and size of the particles. Their results showed that the particles were slow rotators, did not undergo fragmentation, and provided valuable parameters for improving dynamical models of the cometary coma.
Lemos et al. \cite{lemosetal2024} characterized large aggregates ($\geq$1\, cm) from comet 67P. They determined that the aggregates had an average radius of approximately 5\,cm and mapped their source regions near terrain boundaries on the comet's surface, considering the mascon model \cite{Geissleretal96}.
Czechowski \& Kossacki \cite{Czechowskietal2021} examined material ejected from the Hatmehit depression on comet 67P, modeling the ejecta with various initial velocities using the mascon model. Finally, Scheeres \cite{scheeres20122} calculated the location and stability of equilibrium points around comet 67P using the mascon model approach. They found four unstable equilibrium points around comet 67P.

In this work, we analyze in detail the surface dynamics and the orbital dynamics of comet 67P/Churyumov-Gerasimenko from an adequate 3-D polyhedral shape model\footnote{\url{https://pdssbn.astro.umd.edu/holdings/ro-c-multi-5-67p-shape-v2.0/data/triplate/spc_lam_psi/}} obtained by \cite{gaskelletal}. We treat the comet nucleus like the surface of an asteroid, as no drag force was implemented. Our goal is to investigate dynamic features of the surface and surroundings, particularly the slopes, incorporating Third-Body (TB) and Solar Radiation Pressure (SRP) perturbations into our analyses. We utilize an accurate 3-D polyhedral shape model of the nucleus of comet 67P for this purpose. We also compared the Dipole Segment (DS) model to the polyhedron model to identify the limitations for families of planar symmetric periodic orbits around the nucleus of comet 67P, which could support specific orbital phases of space mission planning\footnote{\url{https://www.asteroidmission.org/asteroid-operations/}} \citep{ACCOMAZZO2015434}.

%O que eu tinha colocado:{
% Our goal is to investigate some dynamic features, such as slopes, that may indicate possible regions of accumulation or ejection of material, and to provide an initial theoretical basis of the dynamics of this body that may be of interest for future space missions aimed at exploring small bodies in the Solar System. The importance of studying the surface dynamics and those around comet 67P lies in the fact that it can provide an initial theoretical basis for other future space missions, such as the DESTINY+ mission.
% Such results can help identify more economical trajectories for parking a spacecraft
% in a highly irregular gravitational field. Of course, in the case of a comet similar to 67P, the problems experienced by the Rosetta mission team must be taken into account (e.g., \cite{vallat2017}}

%O que estava antes {
%\textbf{Our goal is to investigate some dynamic features such as slopes, which may indicate possible regions for accumulation or ejection of material, and provide dynamical insights that can be of interest to future space missions aimed at exploring small bodies of the Solar System.} 
%{The importance of studying the surface dynamics and those around comet 67P lies in the fact that it is a peculiar body with a very irregular shape and can be considered a binary, unlike other comets such as comet 1P/Halley or the non-active comet (3200) Phaethon, target of the DESTINY+ \cite{arai2018}.}

% Hence, the importance of trying to model it with analytical models such as the Dipole Segment.

Section \ref{sec:2} describes the methodology used in our analyses. Section \ref{sec:3} presents our results on the surface dynamics of comet 67P, which cover geometric height, surface tilt, geopotential surface, acceleration surface, surface slopes, escape speed, equilibrium points, and zero-velocity curves. Section \ref{sec:4} shows the orbital dynamics around comet 67P from the topological characterization of the families of planar symmetric periodic orbits found around it. We used the DS for this purpose.
Finally, in Section \ref{sec:5}, we present our conclusions.

\section{Methodology}
\label{sec:2}

It is essential to determine the physical characteristics of 67P to calculate and interpret the physical quantities associated with its surface and orbital dynamics \cite{joetal2025}. Modeling the geopotential of small bodies requires strategies that account for their irregular shapes. While methods based on Legendre polynomials offer reasonable approximations, they can introduce discrepancies in certain regions \cite{scheeres1994}. Although computationally efficient, the mass concentration points (mascons) approach, which involves subdividing the body into small mass elements and summing their potentials \cite{Amarante2021}, exhibits significant inaccuracies near the surface. %Alternative models, such as triaxial ellipsoids, attempt to refine gravity field calculations but still have limitations in capturing the detailed gravitational landscape of irregular bodies.}

In this study, we employ the polyhedron method to explore the surface dynamics. Initially developed by \cite{wernner1994,wernerl1996} and later refined by \cite{tsoulis2001, tsoulis2012}, the polyhedron method models the nucleus\footnote{Without loss of generality, we will omit the word \textit{nucleus} along the text.} of comet 67P as a uniform-density polyhedron, enabling analytical computation of its gravitational potential and its derivatives. This approach surpasses alternative models, such as mascons and spherical harmonic expansions, especially when studying regions across and close to the surface \cite{wernerl1996}.

The polyhedron approach is particularly advantageous when studying surface dynamics since it adequately describes gravitational variations caused by the comet's shape. To implement this methodology, we use the Minor-Gravity package\footnote{\url{https://github.com/a-amarante/minor-gravity}} \cite{amaranteandwinter2020, minorgravity}, which computes the gravitational potential of comet 67P based on the 3-D polyhedral shape model and provides its first and second derivatives accurately. We use a current density of 0.535 g\,cm$^{-3}$ inferred by \cite{preusker2015}, from the Rosetta mission, and a rotational period of 12.4043 h derived by \cite{mottola2014}. The polyhedral shape model of comet 67P has 48,420 vertices and 96,834 triangular faces \cite{gaskelletal}. The shape model has a volume of 18.8~km\(^3\) and an equivalent spherical diameter of 3.3~km.
%\textbf{Furthermore, other components of the highly complex force field of Comet 67P's nucleus were not considered for our analysis, such as radiation pressure, perturbations from a third body, or its gas ejection, except the slopes, which were analyzed by adding the gravitational pertubation due to the third-body and also the SRP, both effects analyzed separately}.

In the dynamic orbital analyses, we did not consider other components of the force field. As a result, we treated the comet as an asteroid since we did not implement any drag force. We emphasize that recent work shows that the activity of drag and outgassing in the form of jets when comet 67P is near perihelion may play a fundamental role in the ejection and dynamics of dust aggregates (see \cite{lemosetal2024}). However, its contribution to short-period orbital solutions, far away from the comet surface, such as those investigated in this work, is negligible.
For gravitational perturbations, we can safely assume that the considerable distance of the Sun, compared to the typical size of the close orbits studied by the spacecraft, makes these perturbations negligible. Furthermore, in our surface dynamic analyses, we provided a thorough investigation of how additional forces, such as TB and SRP, affect the overall behavior of slopes on the surface of comet 67P, considering different distances and sizes of particles (sec. \ref{tbsrppert}).

Additionally, to determine and classify equilibrium points around and inside the comet, we apply the Minor-Equilibria-NR package\footnote{\url{https://github.com/a-amarante/minor-equilibria-nr}} \cite{amaranteandwinter2020, minorequilibria}, which computes the location and stability of the equilibria. The Minor-Equilibria-NR package uses the Newton-Raphson 3D method to find equilibria around irregular bodies with the accuracy of 10$^{-5}$.

Finally, we employ the Grid Search Method \citep{Markellosetal1974} with the DS model to compute the families of planar symmetric periodic orbits around comet 67P.
The Grid Search Method is a simple and structured technique that allows the comprehensive identification of sets of families of periodic orbits in non-integrable dynamical systems, considering a delimited area of the space of initial conditions. For the planar case and symmetry about the $x-$axis, it works as follows:
\begin{itemize}[label=--]
    \item A grid $(x_0,J)$ is constructed, since $x_0$ represents the $x$ position at the initial time ($t=0$) and $J=(x^2 + y^2)-2U  -(\dot{x}^2+\dot{y}^2)$ the Jacobi constant. A symmetric periodic orbit crosses the $x-$axis perpendicularly twice, in time $P$ and $P/2$, so $y=\dot{x}=0$, then $J(x_0,0,0,\dot{y}_0)$;
    \item For each set of initial conditions $(x_0, J)$, integrate numerically the equation of motion (\ref{EqMotion}) by TIDES (Taylor series integrator for differential equations), with absolute and relative tolerance equal $10^{-16}$ \citep{abad2012algorithm}, compute the Poincaré map, crossing the $x-$axis and $\dot{y}>0$;
    \item The sign of $\dot{x}$ is checked, and by continuity, if $\dot{x}(t_0+n\Delta t)<0$ and $\dot{x}(t_0+(n+1)\Delta t)>0$, a root-finding process is done \citep{brent1971algorithm};
    \item The initial conditions of the symmetric periodic orbits are satisfied when convergence is achieved;
    \item  Finally, a final check is done by integrating the initial conditions up to the total period $T$. It verifies if $|(x,0,0,\dot{y})-(x(mP),\dot{x}(mP),y(mP),\dot{y}(mP))| < mtol$, being $m$ multiples of periods $P$ and $mtol$ the tolerance. The initial conditions that do not meet the final check are disregarded.
\end{itemize}

%\textbf{We also highlight that an analysis of the relative error between the dipole and the polyhedron model was made and we identified that for distances $>\,\sim$\,3 the dipole model exhibits errors smaller than 3\% compared to the polyhedron model, and can be used, for example, in a first approximation to analyze symmetric and planar periodic orbits around 67P or other small bodies, as long as we consider an idealized model, without activity on its surface, gas jet forces, for example}.
% This methodological framework ensures a characterization of the geopotential of comet 67P, supporting further investigations into its surface dynamics and orbital environment.

\section{Surface Characteristics}
\label{sec:3}
\subsection{Geometric Height}
\label{sec3:1}
\begin{figure}[h]
\centering
\includegraphics[width=0.6\textwidth]{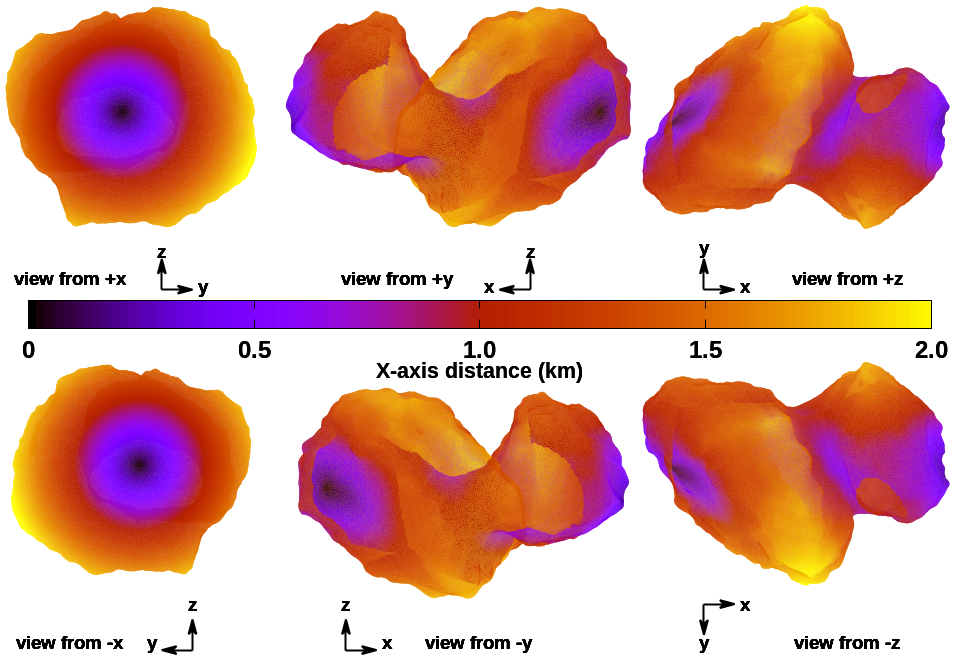}
\caption{3-D polyhedral shape model in 3-D of 67P shown in six perspective views ($\pm$x, $\pm$y, and $\pm$z). The shape model was built with 48,420 vertices and 96,834 triangular faces. The color code indicates the distance from the centroid facet to the $x-$axis in kilometers (geometric height).}
\label{fig:1}
\end{figure}

Geometric height is defined as the distance between the centroid of each triangular face on the surface of a body and the main $x-$axis \cite{amaranteandwinter2020}. This feature, along with the surface's tilt angles (see Section \ref{sec:3:2}), is a tool for analyzing the topography of elongated and binary minor celestial bodies,such as comet 67P. Figure~\ref{fig:1} reproduces the 3-D polyhedral shape model of comet 67P, using 48,420 vertices and 96,834 triangular faces \cite{gaskelletal}. The color scale indicates the geometric height. The colors range from purple to yellow, highlighting the shapes of the small and big lobes of comet 67P. Furthermore, El-Maarry \cite{elmarry} represented the topographic features of comet 67P. As we can see, the geometric height ranges from zero (purple areas in Fig.~\ref{fig:1}, such as Ma'at, Anuket, and Maftet, named in Fig.~2 from \cite{elmarry}) to 2.0~km (yellow areas in Fig.~\ref{fig:1}, such as Atum, Bes, and Anhur, named in Fig.~2 from \cite{elmarry}).
\subsection{Surface Tilt}
\label{sec:3:2}
\begin{figure}[H]
\centering
\includegraphics[width=0.7\textwidth]{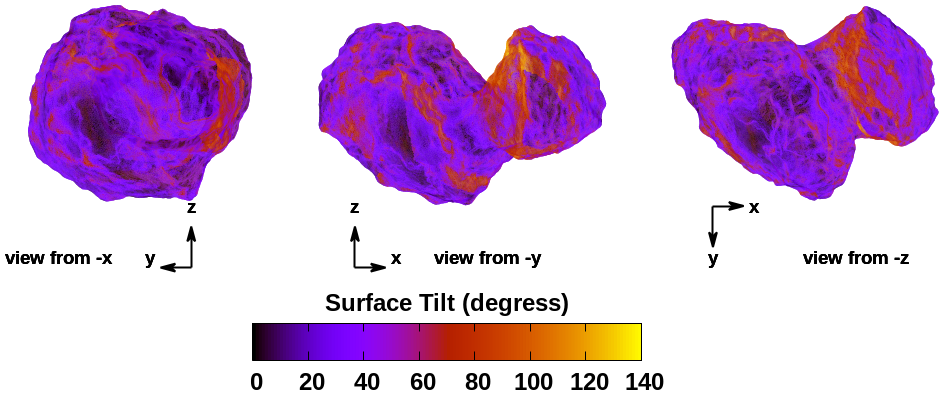}
\caption {Surface tilt is mapped across the comet 67P, considering the 3-D polyhedral shape model with 48,420 vertices and 96,834 faces. The color code gives the surface tilt in degrees.}
\label{fig:2}
\end{figure}

Let be the vector that starts from the center of mass of 67P and points to a given location $\mathbf{r}$ on its surface. The angle between the normal surface vector $\mathbf{\hat{n}}$ and the barycenter vector $\mathbf{r}$ is called the tilt angle \cite{scheeresetal2016}. This angle helps to understand how each triangular face is oriented within the comet's shape. Fig. \ref{fig:2} shows the mapping of the tilt across the surface of comet 67P. The tilt values ranges from 0$^{\circ}$ to 140$^{\circ}$, with higher values over the small lobe, reaching $\sim135^{\circ}$. Our analysis also suggests that 98.5\% of its surface has tilt values that do not exceed 100$^{\circ}$.
\subsection{Geopotential Surface}
\label{sec:3:3}
The geopotential surface (the sum of gravitational potential and centrifugal potential) of comet 67P, with a polyhedral shape model at its center of mass and aligned with the principal axes of inertia, is given by \cite{scheeres2012}.
\begin{equation}
V(\mathbf{r}) = -\frac{1}{2} \omega^2 \left(x^2 + y^2\right) + U(\mathbf{r}), 
\label{eq:1}
\end{equation}
\noindent where $\mathbf{r}$ represents the position of the particle in the body-fixed reference frame relative to the center of mass, $\omega$ is the angular speed, and $U(\mathbf{r})$ is the gravitational potential computed numerically \cite{tsoulis2001, tsoulis2012} or analytically (Eq.~\eqref{potDS}).

\begin{figure}[H]
\centering
\includegraphics[width=0.8\textwidth]{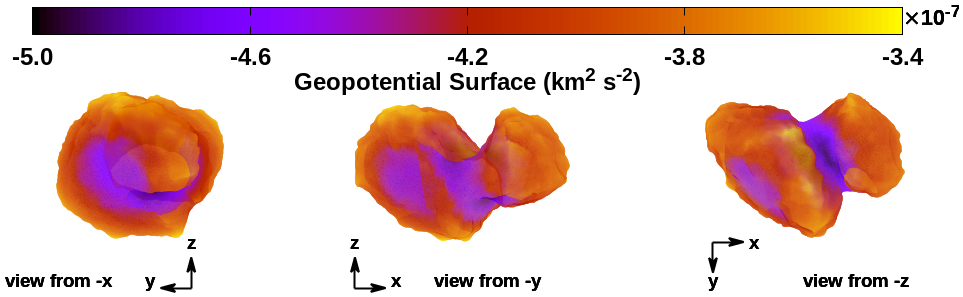}
\caption{Map of the geopotential computed across the surface of the comet 67P. The color bar gives the numerical values of Eq. \eqref{eq:1}, in km${^2}$s$^{-2}$.}
\label{fig:3}
\end{figure}

Fig. \ref{fig:3} shows the geopotential computed over the surface of comet 67P. Due to its slow rotation, the gravitational potential predominates. We observed that the regions with the minimum geopotential values are located in the Hapi region (comet 67P's neck) and the big lobe. Comparing the values of the geopotential of comet 67P with those found on the surface of the Kuiper Belt Object (KBO), Arrokoth \cite{amaranteandwinter2020}, that has an irregular binary shape with similar density and rotational period of comet 67P. We can see that comet 67P has a minimum geopotential $\sim20\times$ lower than KBO Arrokoth. %Additionally, the Hapi region of comet 67P exhibits the lowest geopotential values.% We can see that comet 67P has a geopotential $\ sim$0.0425$\times$ greater than Arrokoth. Additionally, the Hapi (neck) region is characterized by the minimum geopotential values.
% -10/(-5*10^(-7)*(1000)^2) = 20
\subsection{Surface Acceleration}
\label{sec:3:4}
Given the geopotential at any location \( \mathbf{r} \) on the surface of comet 67P, by calculating the gradient:
\begin{align}
 \|-\nabla V(\mathbf{r})\|,
 \label{eq:2}
\end{align} we obtain the surface acceleration about comet 67P.

\begin{figure}[H]
\centering
\includegraphics[width=0.8\textwidth]{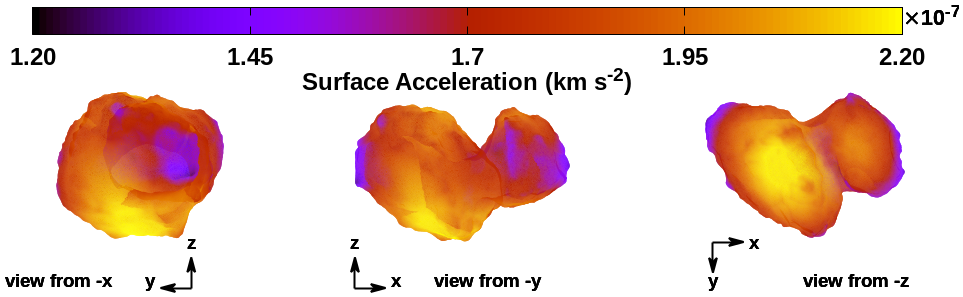}
\caption{Surface acceleration computed over the surface of comet 67P showed in three perspective views (-x, -y, and -z). The color code gives the surface acceleration, in km\,s$^ {-2}$.}
\label{fig:4}
\end{figure}

We calculate the surface acceleration as the sum of gravitational and centrifugal accelerations. Its variation can affect the movement of loose material on the surface of comet 67P and help predict possible preferential regions for material accumulation associated with surface slopes (see Section \ref{sec:3:6}). Fig. \ref{fig:4} shows the surface acceleration across the surface of comet 67P. The surface acceleration was calculated at the barycenter of each triangular face of the 3-D polyhedral shape model of the comet 67P. The color bar shows the surface acceleration.

The big lobe has the maximum values of surface acceleration, which corroborates the minimum values of geopotential. In contrast, the Hapi region (comet 67P neck) has intermediary values for the surface acceleration. This behavior indicates that the gravitational potential dominates in the large lobe. Moreover, the centrifugal potential has more influence in the neck region, which is the distance that is much closer to the center of mass. The surface acceleration reaches $\sim2.2\times10^{-7}$km\,s$^{-2}$. Comparing the results of the maximum surface acceleration with those found for KBO Arrokoth \cite{amaranteandwinter2020}, we observe that comet 67P has values $\sim4\times$ lower than Arrokoth.%we observe that comet 67P has surface acceleration values $\sim$0.231$\times$ lower than Arrokoth.
% 0.95/(2.2*10^(-7)*1000000) = 4.31818181818182
\subsection{Escape Speed}
\label{sec:3:5}
The local normal escape speed $v_{es}$ was computed across the surface of comet 67P using Eq. \eqref{eq:3} \cite{scheeres2012}.
\begin{equation}
v_{es} = -\mathbf{\hat{n}} \cdot (\boldsymbol{\Omega} \times \mathbf{r}) + 
\sqrt{\left[ \mathbf{\hat{n}} \cdot (\boldsymbol{\Omega} \times \mathbf{r}) \right]^2 - 2U_{{\text{min}}} - (\boldsymbol{\Omega} \times \mathbf{r})^2},
\label{eq:3}
\end{equation}
\noindent where $\mathbf{r}$ is the radius vector from the comet's center of mass to the local surface, \( U_{\text{min}} = \min \left[ U(\mathbf{r}), -\frac{GM}{\|\mathbf{r}\|} \right] \), $G = 6.67428\times10^{-20}$\,km$^3$\,kg$^{-1}$\,s$^{-2}$ is the gravitational constant and $M = 1.0\times10^{13}$\,kg is the total mass of comet 67P and \(\Omega\) is the angular velocity vector.

\begin{figure}[H]
\centering
\includegraphics[width=0.8\textwidth]{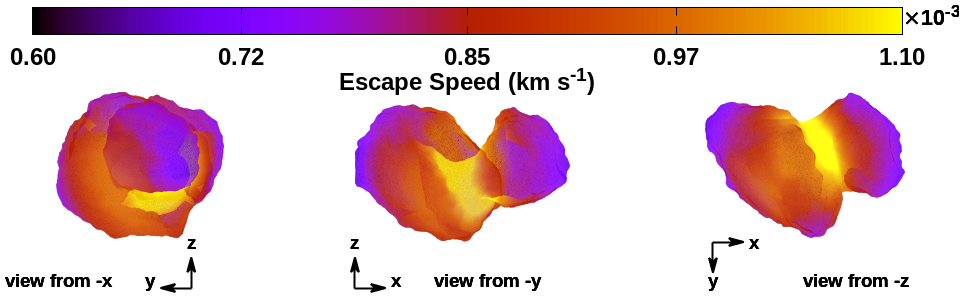}
\caption{Local normal escape speed calculated over the surface of the comet 67P, in km\,s$^{-1}$. The color box denotes the values of escape speed.}
\label{fig:5}
\end{figure}

If the launch speed exceeds the given speed, the particle will escape. Our approach is based on the assumption that the particle is launched normally to the local surface. Fig. \ref{fig:5} shows the escape speed computed over comet 67P's surface. 
%\textbf{(The escape speed has higher values in the Hapi region (neck). This behavior is similar across the surface of minor bodies with contact binary shapes, such as KBO Arrokoth \cite{amaranteandwinter2020}}).
The regions of the minimum geopotential are the areas with the maximum escape speed. 
We observed that the Hapi region has the maximum escape speed. This behavior is similar to the minor bodies with slow rotational periods, such as Arrokoth, Toutatis, Leucus, and Justitia, for example. The escape speed values in the neck of comet 67P can reach $1.10\times10^{-3}$km\,s$^{-1}$.
%\textbf{(The regions of higher values from the escape speed are approximately the locations of minimum geopotential (Fig. \ref{fig:3}) because comet 67P rotates slowly.)}
%Comparing with the Arrokoth, we observe that the intensity of accelerations in the neck of 67P is $\sim$0.129$\times$ lower than that of Arrokoth.
%We can see that the Hapi region has the maximum values of escape speed similarly to what happens with the geopotential (Fig. \ref{fig:3})
\subsection{Surface Slopes}
\label{sec:3:6}
Considering the surface acceleration vector calculated on the barycenter of each triangular face of the comet 67P, using Eq. \eqref{eq:2}, let us analyze how each face is oriented about this vector.
Let be the angle between the normal unit vector $\hat{n}$ on the surface and the surface acceleration vector $-\nabla V(\mathbf{r})$ of a given face. The slope $\theta$ is defined as the supplement of this angle \cite{scheeres2012, scheeresetal2016}.
 \begin{align}
%\theta = 180^\circ - \arccos\left(\frac{-\nabla V(\mathbf{r}) \cdot \mathbf{\hat{n}}}{\| -\nabla V(\mathbf{r}) \|}\right)\dfrac{180^{\circ}}{\pi}.
\theta = \dfrac{180^\circ}{\pi}\bigg[1 - \arccos\left(\frac{-\nabla V(\mathbf{r}) \cdot \mathbf{\hat{n}}}{\| -\nabla V(\mathbf{r}) \|}\right)\bigg].
\label{eq:4}
\end{align}
This dynamic characteristic provides insight into the movement of particles across the surface of comet 67P. If $\theta>90^{\circ}$, there is a possibility that the particles will be ejected from the surface of comet 67P, while for $\theta<90^\circ$, the particles could accumulate in some regions over the comet's surface.

Using the mascon approach, Scheeres \cite{scheeres20122} analyzed the surface slopes across the surface of comet 67P. They found that the surface slopes range from nearly $0^\circ$ in the smoothest regions to around $30^\circ$ in the steepest areas. %, indicating a variation from flat terrains to slopes approaching the angle of repose of surface materials.
\cite{Czechowskietal2021}, also using the mascon approach, found that most values of surface slopes are between $20^\circ-40^\circ$ and $60^\circ-90^\circ$. The range $20^{\circ}-40^{\circ}$, accounting for $\sim30$\% of the comet 67P surface, corresponds to moderately inclined terrains, which may play a crucial role in mass-wasting processes.
Meanwhile, the range $60^{\circ}-90^{\circ}$ covers a small portion ($\sim11$\%), representing the steepest regions, likely associated with cliff faces or highly eroded areas, where material instability is more pronounced.
Finally, Vincent \cite{vicentetal2017} uses the polyhedral method to calculate the slopes ($\theta>60^\circ$) to indicate erosional age.% They focus on $\theta>60^\circ$ to isolate cliffs and abrupt features and correlate their distribution with the erosion processes acting on the cometary surface. 

\begin{figure}
\centering
\includegraphics[width=0.7\textwidth]{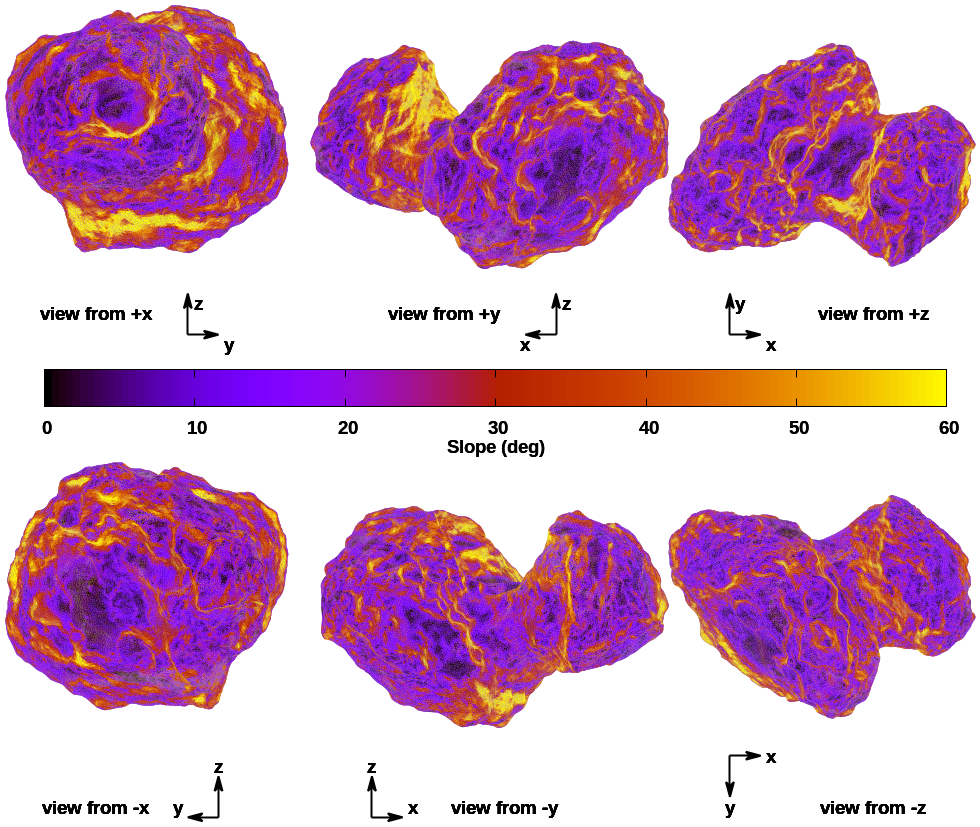}
\caption{Surface slopes mapped across the surface of comet 67P viewed in six perspectives ($\pm$x, $\pm$y, and $\pm$z). The color bar indicates the angle $\theta$ (degrees) computed numerically from Eq. \eqref{eq:4} over the surface of comet 67P.}
\label{fig:6}
\end{figure}

We can see from Fig. \ref{fig:6} that comet 67P presents regions with low values of $\theta$. However, we can observe some areas with high values of surface slopes ($\theta > 140^{\circ}$). This behavior suggests that a few regions on the surface of comet 67P may be ejecting particles.
Table \ref{tab:1} presents our statistics on the surface slopes of comet 67P, utilizing a suitable 3-D polyhedral shape model, with current density of 0.535 g\,cm$^{-3}$ from the Rosetta mission, and a rotational period of 12.4043 h \cite{preusker2015,mottola2014}.
Compared with previous works, the most significant differences are between intervals $20^\circ-40^\circ$ and $60^\circ-90^\circ$. From Tab. 3 of \cite{Czechowskietal2021}, the surface slopes between $20^\circ-40^\circ$ represent only 30\% of the total surface area of comet 67P, while in Tab. \ref{tab:1} this represents 38\%. Meanwhile, the surface slopes in the interval of $60^\circ-90^\circ$ cover 11\% of the surface area of comet 67P. Tab. \ref{tab:1} indicates a coverage of only $\sim$3\%.
\begin{table}
\centering
\caption{Distribution (\%) of surface slopes $\theta$ (degrees) on the comet 67P, considering a shape model of 96,834 faces and 48,420 vertices \cite{gaskelletal}.}
\renewcommand{\arraystretch}{1.1}
\setlength{\tabcolsep}{28pt} 
\label{tab:1}
\begin{tabular}{ccc}
\hline
Slope Range (degrees) & Number of Faces & Percentage (\%) \\
\hline
0 -- 20  & 42,853 & 44.25\\
20 -- 40 & 38,093 & 39.34\\
40 -- 60 & 13,132 & 13.56\\
60 -- 90 & 2,753   & 2.84 \\
90 -- 180 & 3 &     0.01 \\
\hline
Total & 96,834 & 100 \\
\hline
\end{tabular}
\end{table}
\subsection{Equilibrium Points and Zero-Velocity Curves}
\label{sec:3:7}
\begin{figure}
\centering
\includegraphics[width=0.7\textwidth]{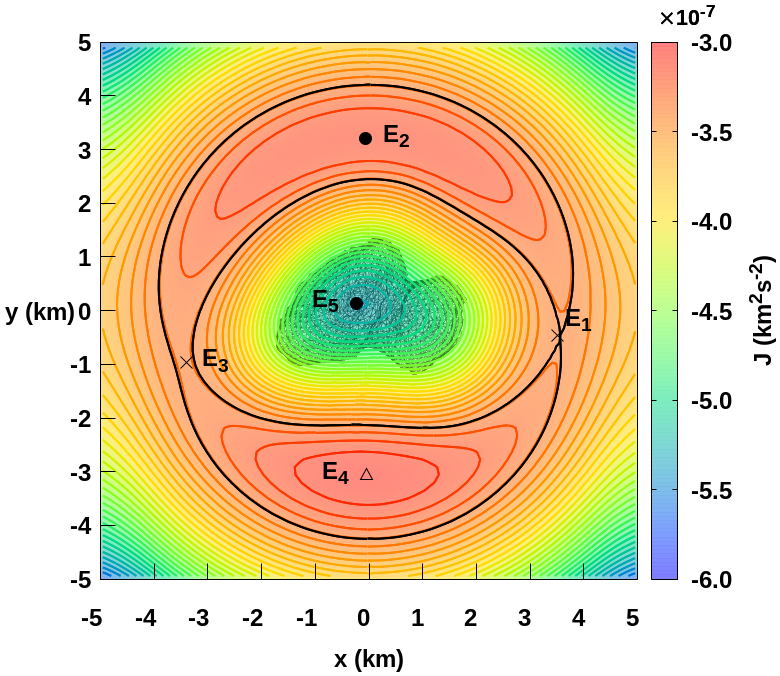}
\caption{Location of the five equilibrium points around comet 67P in the $xOy$ projection plane. Equilibrium points denoted by crosses are topologically classified as Case II (hyperbolically unstable), equilibrium points with circles are referred to as Case I (linearly stable), and equilibrium points with a triangle are denominated as Case V (complexly unstable). The region delimited by the black line intersecting the equilibrium point E$_1$ represents the rotational Roche lobe of comet 67P.}
\label{fig:8}
\end{figure}

%Studying orbits around small celestial bodies has become increasingly relevant, particularly for space exploration and sample return missions. Therefore, it is interesting for a space mission to determine the location and stability of equilibrium points. These findings are essential for understanding the behavior of particles around comet 67P, providing important insights into its dynamics and potential interactions with future space missions.
In this section, we calculate the location, number, and stability of the external and internal equilibrium points of comet 67P, using Eq. \eqref{eq:5}.
\begin{equation}
\nabla V(\mathbf{r}) = 0.
\label{eq:5}
\end{equation}
We identified five equilibrium points based on an adequate 3-D polyhedral shape model of comet 67P, which comprises 96,834 faces and 48,420 vertices \cite{gaskelletal}. Figure \ref{fig:8} shows the location of the five equilibrium points with different topological characteristics viewed in the $xOy$ plane. The equilibrium points E$_1$ and E$_3$ are unstable and classified as Case II \cite{jiang2015}. Equilibrium point E$_4$ is also unstable and classified as Case V. Unlike Scheeres \cite{scheeres20122}, which identified four equilibrium points around comet 67P, all of them unstable, the equilibrium points E$_2$ and E$_5$ were linearly stable and classified as Case I, indicating that particles surrounding these equilibrium points may have timescales greater than the unstable equilibrium points E$_1$, E$_3$, and E$_4$. These equilibrium point regions may offer reduced fuel costs under idealized conditions.
Additionally, equilibrium point E$_5$ is located inside the comet 67P close to its center of mass.

%\textbf{These results suggest that these regions may offer reduced fuel costs under idealized conditions (e.g., around equilibrium point E$_2$) or even for remaining particles in a synchronic orbit that are ejected from the surface of comet 67P. }

% It is worth highlighting that ESA's Rosetta mission successfully landed the space probe on the comet's surface in September 2016, reinforcing the importance of more realistic and refined models for predicting orbital dynamics in such complex environments, \textbf{since \cite{vallat2017} analyses demonstrate the difficulties faced by the probe to land on its surface, mainly due to the ejection of its gas.}

Furthermore, Tab. \ref{tab:2} shows the location of the equilibrium points of comet 67P. In addition, we indicate their latitude $ \varphi $ (degrees), longitude $\lambda $ (degrees), radial distance $r$ from the center of mass (km), and geopotential $V(\mathbf{r})$ (km${^2}$s$^{-2}$). The colored contour lines indicate the Jacobi constant (J) values associated with zero-velocity curves, in (km${^2}$s$^{-2}$). The zero-velocity curve behavior confirms that 67P has a total of five equilibrium points.

The equilibrium points within the body’s rotational Roche lobe limit (Fig. \ref{fig:8}) may eventually trap particles \cite{scheeresetal2019, Amarante2021}. Our results show that the rotational Roche lobe, the teardrop-shaped region surrounding comet 67P, limited by the black line in Fig. \ref{fig:8}, intersects itself at the E$_1$ equilibrium point, while in \cite{scheeres20122}, it intersects itself at the equilibrium point E$_3$. This feature indicates that equilibrium point E$_1$ has the minimum value for the geopotential, which differs from previous works \cite{scheeres20122}, which found the minimum of the geopotential at equilibrium point E$_3$.
\begin{table}
\caption{Location of equilibrium points around comet 67P and their geopotential values. Equilibria are computed through the polyhedron method with the Minor-Equilibria package \cite{amaranteandwinter2020} and an accuracy of \(10^{-5}\).}
\label{tab:2}
\renewcommand{\arraystretch}{1.1}
\setlength{\tabcolsep}{5pt} 
\begin{tabular*}{\textwidth}{@{\extracolsep{\fill}}lccccccc}
\toprule
Point & $X$ (km) & $Y$ (km) & $Z$ (km) & $\varphi$ (°) & $\lambda$ (°) & $r$ (km) & $V(\mathbf{r})$ ($\times10^{-7}$km$^{2}$s$^{-2}$)   \\
\midrule
E$_1$ & 3.50342 & -0.47082 & 0.09106 & 1.475 & 352.346 & 3.53541 & -3.30661 \\
E$_2$ & -0.06365 & 3.20769 & -0.05010 & 359.105 &  91.136 & 3.20831 & -3.08010 \\
E$_3$ & -3.39552 & -0.96706 & 0.06879 & 1.116 & 195.897 & 3.53801 & -3.29503 \\
E$_4$ & -0.05399 & -3.06035 & -0.03567 & 359.933 & 268.990 & 3.06103 & -3.01194 \\
E$_5$ & -0.23159 & 0.13858 & -0.14763 & 331.320 & 149.102 & - & -5.58168\\
\bottomrule
\end{tabular*}
\end{table}

The equilibrium points pair E$_1$-E$_5$ has a large $z$-component ($\sim 10^{-1}$) when compared with the others equilibrium points ($\sim 10^{-2}$). Table \ref{tab:2} also illustrates that the radial distance to the center of mass highlights an axial symmetry that occurs particularly between the equilibrium point pairs E$_1$–E$_3$ and E$_2$–E$_4$, which are located on opposite sides of the comet 67P. Furthermore, the equilibrium points E$_1$ and E$_3$ lie above the $xOy$ plane, while the equilibrium points E$_2$, E$_4$, and E$_5$ remain below it.

\section{\textbf{Third-Body and Solar Radiation Pressure Perturbations}}
\label{tbsrppert}
\subsection{Third-Body Gravitational Perturbation}
In this section, we analyzed the slopes over the surface of 67P considering the gravitational perturbation of a Third-Body (Jupiter (J) or Sun (S), J/S) at a particle located at the barycenter of each face of the 3-D polyhedral shape model, using Eq. \eqref{eqofmotionpertubed}:

\begin{equation}
\mathbf{a_{{pert}}}(\mathbf{r}) = G M_{\text{J/S}} \left(
\frac{\mathbf{r}_{\text{J/S}} - \mathbf{r}}{\|\mathbf{r}_{\text{J/S}} - \mathbf{r}\|^3} - 
\frac{\mathbf{r}_{\text{J/S}}}{\|\mathbf{r}_{\text{J/S}}\|^3}
\right),
\label{eqofmotionpertubed}
\end{equation}
\noindent where $M_\text{J} = 1.9\times10^{27}$\,kg is the total mass of Jupiter, $M_\text{S} = 2.0\times10^{30}$\,kg is the total mass of Sun, $\mathbf{r_J}$ is the position vector of Jupiter concerning 67P, $\mathbf{r_S}$ is the position vector of the Sun concerning 67P, $\|\mathbf{r_J-r}\|$ is the distance from the barycenter vector of each face to Jupiter, and $\|\mathbf{r_S-r}\|$ is the distance from the barycenter vector of each face to the Sun.

The slopes $\theta$ now could be computed from Eq. \eqref{eq:theta}:

\begin{equation}
\theta = \dfrac{180^\circ}{\pi}\bigg[1 - \arccos\left(\frac{-\mathbf{a}_{\mathbf{{total}}}(\mathbf{r}) \cdot \mathbf{\hat{n}}}{\| -\mathbf{a}_{\mathbf{{total}}}(\mathbf{r}) \|}\right)\bigg],
\label{eq:theta}
\end{equation}

\noindent where $\mathbf{{a}_{{total}}=a_{pert}{(r)-\nabla V(\mathbf{r})}}$, and $\mathbf{\hat{n}}$ is the normal vector calculated at the centroid of each face.

\begin{figure}
\centering
    \subfloat[]{\includegraphics[width=0.45\textwidth]{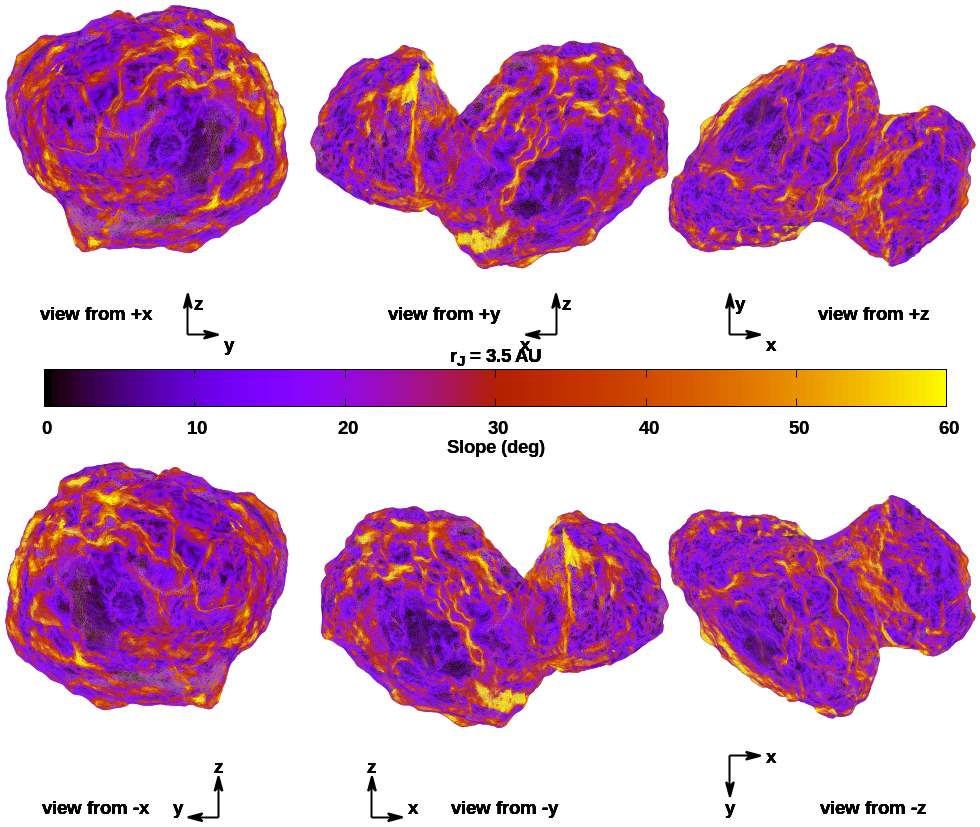}}
    \hspace{0.5em}
    \subfloat[]{\includegraphics[width=0.45\textwidth]{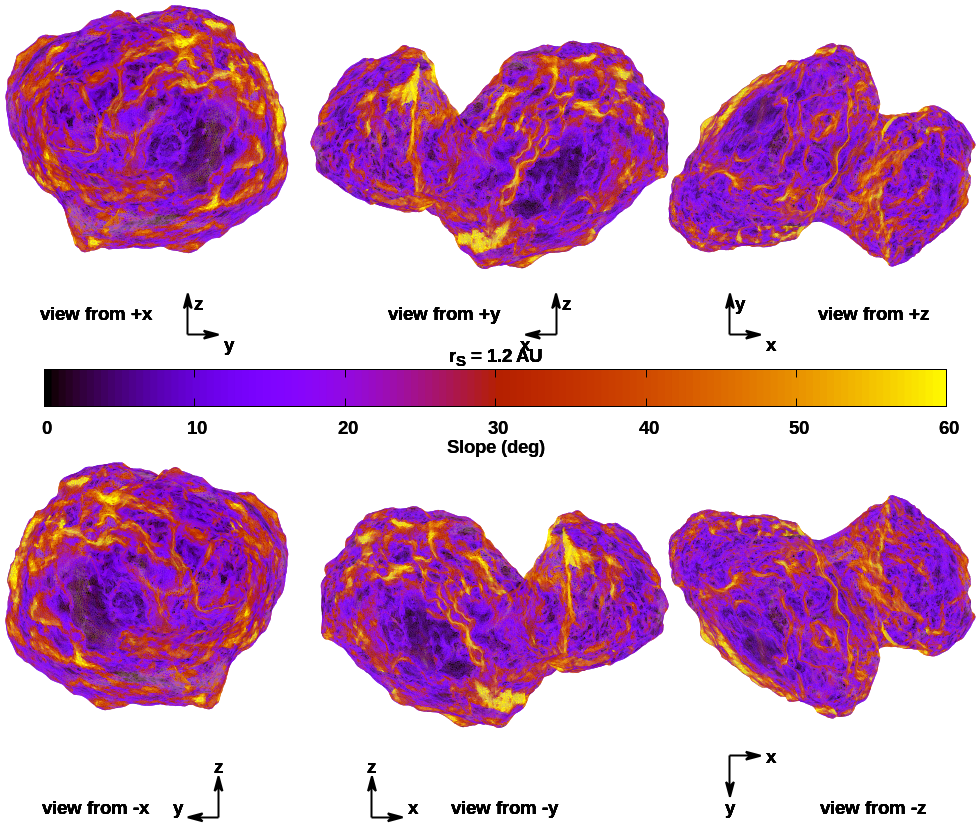}}
    \caption{Surface slopes mapped on the surface of comet 67P seen from six perspectives ($\pm$x, $\pm$y, and $\pm$z) considering the Jupiter (a) and Sun (b) gravitational perturbations at 3.5 AU and 1.2 AU distance from comet 67P, respectively.}
    \label{jupsun}
\end{figure}

Figure \ref{jupsun} shows the global behavior of the slopes due to the gravitational effects of the TB (Jupiter/Sun). We can observe that the TB does not influence significantly the slopes over the comet's surface, when we take into account only the gravitational acceleration from comet 67P (Fig. \ref{fig:6}), which corroborates with the Hill radius of comet 67P, being 318\,km \citep{Rtound2015}.

% \textbf{Figure \ref{pertubedslopesun} shows the variation of the slopes due to the gravitational effects of the third-body (Sun), the analyses remain the same as in the previous figure (see Fig. \ref{pertubedslopejupiter})}.
\subsection{Solar Radiation Pressure Perturbation}
Another force considered for the slope analysis was the Solar Radiation Pressure (SRP). For our analyses, we once again calculated the effect of SRP on a particle located in each face of the polyhedral shape model, assuming that the particle is located at the barycenter of each face.
The equation of motion for a particle with geometric cross-section $A$ and mass $m$ moving under the influence of the SRP acceleration is given by Eq. \eqref{eqofsrp} \citep{Burns1979, Mingard1984, Amarante2021}:
\begin{equation}
\mathbf{a}_{\mathrm{SRP}}(\mathbf{r}) = - \frac{Q_{\mathrm{pr}} S_{0} R_{0}^{2}}{c \, \|\mathbf{r}_{s} - \mathbf{r} \|^{3}} \, \frac{A}{m} \, (\mathbf{r}_{s} - \mathbf{r}),
\label{eqofsrp}
\end{equation}
\noindent where $\mathbf{r}_S$ is the position vector of the Sun relative to 67P, $\|\mathbf{r}_S - \mathbf{r}\|$ is the distance from the particle's located at the barycenter of each face to the Sun, $S_0$ is a solar constant or radiation flux density at the distance of the astronomical unit ${R_0}$ (AU), $c$ is the speed of light, and $Q_{pr}$ is the dimensionless efficiency factor for radiation pressure, which depends on the properties (e.g., density, shape, size) of the particle. In the mathematical model, this value is assumed to be $1$ to represent the value of an ideal material. The values of the adopted SRP parameters are presented in Tab.~\ref{tab:srp_params}.
Furthermore, the particles in the mathematical model were assumed to be fully spherical, in order to vary the area-to-mass ratio, as shown in Eq. \eqref{areamassratio}:
\begin{equation}
    \frac{A}{m}=\frac{3}{4}\frac{1}{\rho r_p},
\label{areamassratio}
\end{equation}
\noindent where $\mathbf{\rho}$ is the particle bulk density, assumed to be equal to the comet 67P-averaged bulk density for all particles, and $\mathbf{r_p}$ is the particle size (spherical radius) (Tab. \ref{tab:srp_params}).
\begin{table}
\centering
\caption{The Solar Radiation Pressure parameters used in the mathematical model.}
\begin{tabular}{l l l l}
\toprule
Parameter & Value & Units & Comments \\
\midrule
$Q_{pr}$ & ${1}$ & $--$ & Efficiency factor \\
$S_0$ & $1.36 \times 10^3$ & kg\,s$^{-3}$ & Solar constant \\
$R_0$ & $1.495978707 \times 10^8$ & km & Astronomical unit distance (AU) \\
${c}$ & $2.99792 \times 10^5$ & km\,s$^{-1}$ & Speed of light \\
$\rho$ & $1.190$ & g\,cm$^{-3}$ & density \\
$r_p$ & $10^{-4}-10^{-1}$ & cm & Particle sizes (spherical radius) \\
\bottomrule
\end{tabular}
\label{tab:srp_params}
\end{table}

From this, we obtain the SRP acceleration vector for a particle located in each face of the 3-D polyhedral shape model and then add it to the acceleration vector used in section \ref{sec:3:6}, considering only the gravitational acceleration from the comet 67P. Thus, we obtain the new total acceleration vector due to SRP, given in Eq. \eqref{totalsrpacc}.
\begin{equation}
\mathbf{a_{total}} = \mathbf{a_{SRP}}(\mathbf{r}) {-}{\nabla \mathbf{V}(\mathbf{r})}.
\label{totalsrpacc}
\end{equation}
\noindent Then, we analyze the global behavior of the slopes, varying the particle size and the distance from the Sun, at pericenter $r_{pe}=1.2$ (AU) and apocenter $r_{ap}=5.7$ (AU) of the comet 67P orbit\footnote{\url{https://ssd.jpl.nasa.gov/tools/sbdb_lookup.html\#/?sstr=Churyumov-Gerasimenko}}.

\begin{figure}
\centering
    \subfloat[Particle size 10$^{-4}$\,cm at pericenter $r_{pe}=1.2$ (AU).]{\includegraphics[width=0.45\textwidth]{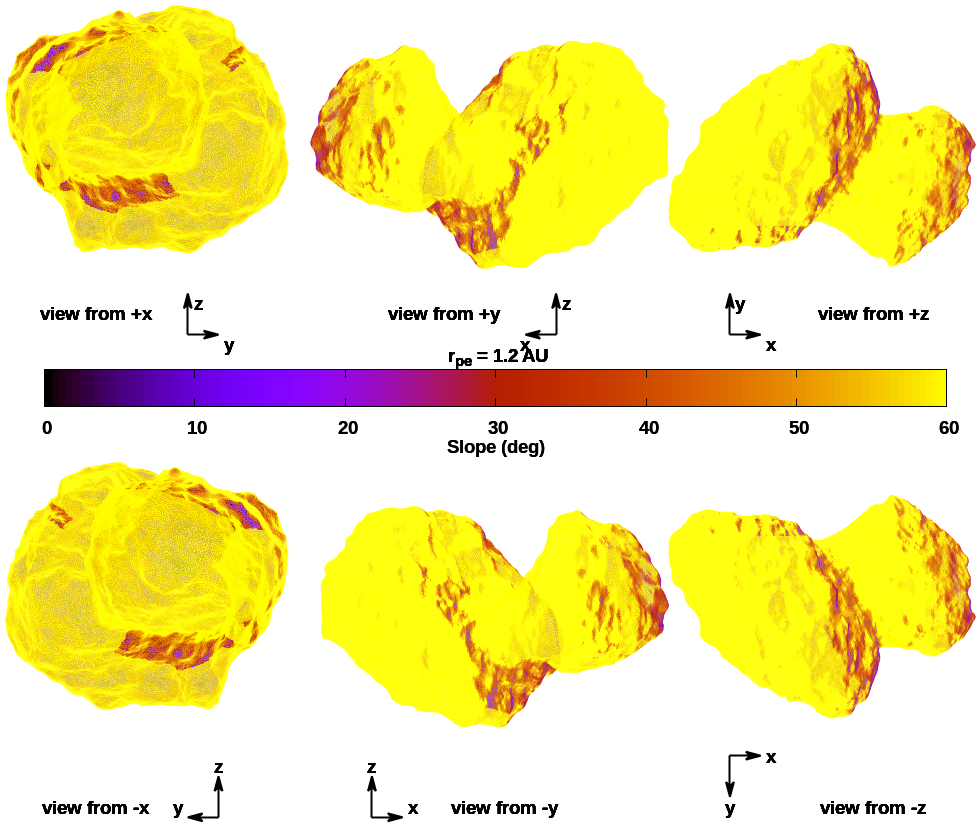}}
    \hspace{0.5em} % Espaço reduzido entre as figuras
    \subfloat[Particle size 10$^{-1}$\,cm at pericenter $r_{pe}=1.2$ (AU).]{\includegraphics[width=0.45\textwidth]{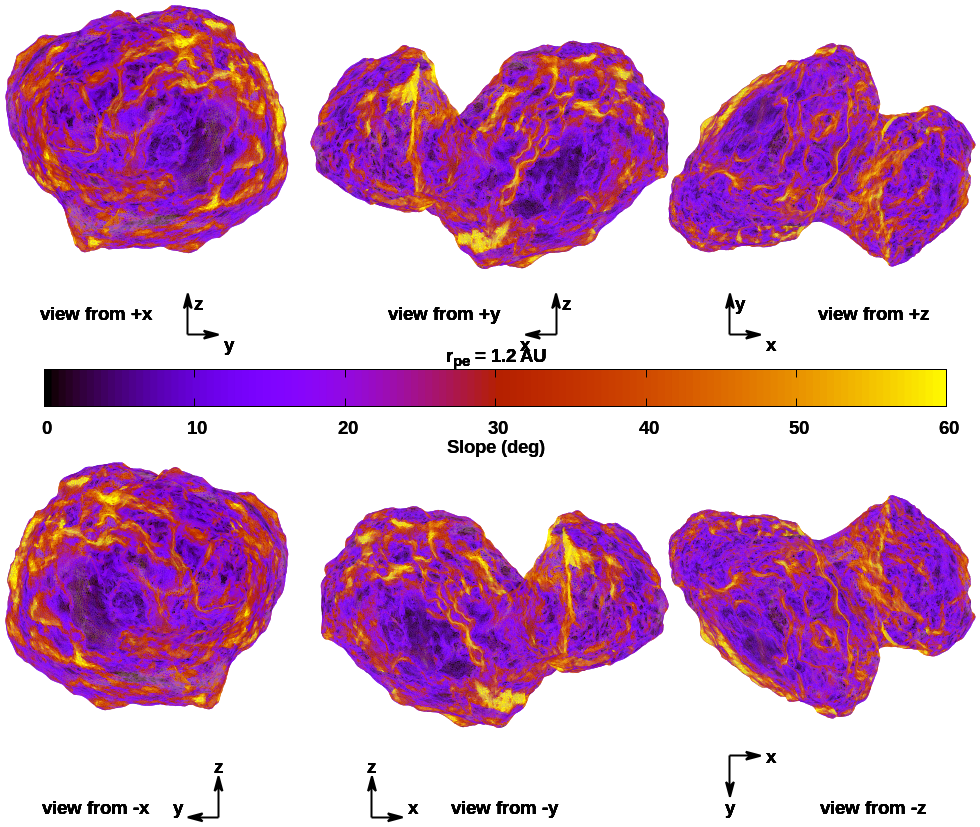}}
    
    \vspace{1em} % Espaço vertical entre as linhas
    
    \subfloat[Particle size 10$^{-3}$\,cm at apocenter $r_{ap}=5.7$ (AU).]{\includegraphics[width=0.45\textwidth]{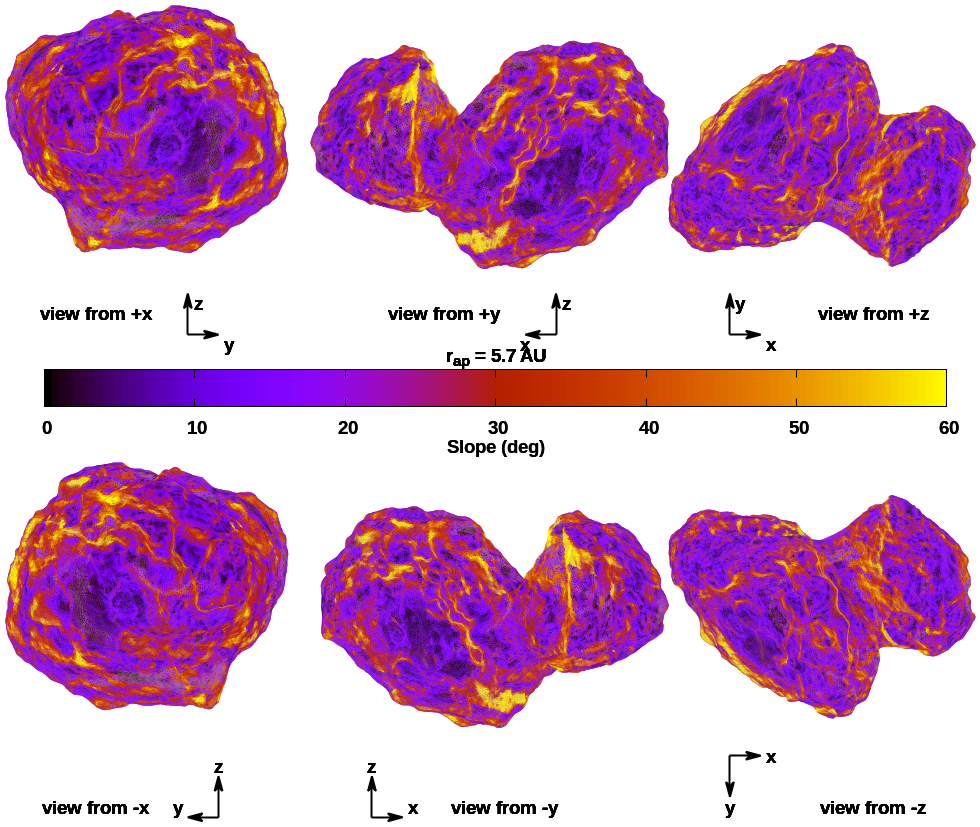}}
    \hspace{0.5em} % Espaço reduzido entre as figuras
    \subfloat[Particle size 10$^{-3}$\,cm at pericenter $r_{pe}=1.2$ (AU).]{\includegraphics[width=0.45\textwidth]{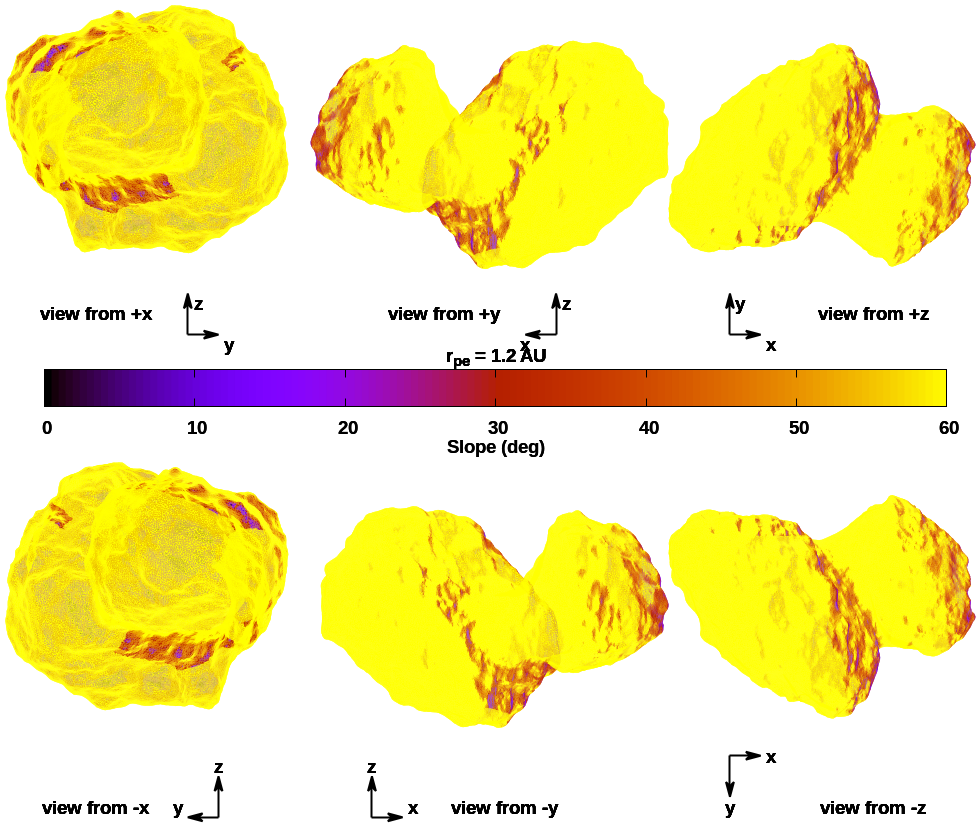}}
    
    \caption{(a) and (b): Global behavior of slopes as a function of particle sizes, keeping the comet distance from the Sun. (c) and (d): Global behavior of slopes as a function of comet distance from the Sun, keeping the particle size. The color scale shows the slopes from 0$^\circ$ to 60$^\circ$.}
    \label{apo_peri_rp}
\end{figure}

In Figs. \ref{apo_peri_rp} (a) and (b), we kept the pericenter $r_{pe}=1.2$ (AU) of comet 67P's orbit and varied the particle sizes from $10^{-4}$\,cm to $10^{-1}$\,cm, located at the barycenter of each face of the 3-D polyhedral model. We found that particles $<\sim10^{-4}$\,cm are significantly influenced by SRP at the pericenter of comet 67P's orbit, while particles on the order of $\sim10^{-1}$\,cm are not significantly influenced by SRP.

In Figs. \ref{apo_peri_rp} (c) and (d) we kept the size of the particles at $10^{-3}$\,cm, located at the barycenter of each face of comet 67P, and varied its distance from the Sun at pericenter $r_{pe}=1.2$ (AU) and apocenter $r_{ap}=5.7$ (AU) of its orbit. We identify that particles with sizes of $10^{-3}$\,cm are significantly influenced by the SRP at the pericenter. In contrast, particles of the same size do not significantly suffer the influence of the SRP at the apocenter.

It is important to note that the acceleration values due to SRP were calculated based on an assumed area exposed to solar radiation of approximately 64\,m$^2$ \citep{ACCOMAZZO2015434}, which represents the maximum contribution of the SRP. So, we can safely neglect the SRP effects in the analyses of orbital dynamics of a spacecraft around comet 67P (sec. \ref{sec:4}).
\subsection{Comparison Between Dipole Segment and Polyhedron Models} 
In this section, we use the Eq. \eqref{relativeerroreq} to calculate the relative error between the gravitational potentials of the polyhedron and the Dipole Segment (DS) models in percentage.
\begin{equation}
\left|\frac{  U_{DS}(x,y)-U_{p}(x,y)}{U_{p}(x,y)}\right|,
\label{relativeerroreq}
\end{equation}
\noindent where $U_{DS}(x,y)$ is the gravitational potential from the DS model and $U_{p}(x,y)$ is the gravitational potential from the polyhedron model, both computed in points around comet 67P, in the $xOy$ plane.

\begin{figure}
    \centering
    \includegraphics[width=1.0\linewidth]{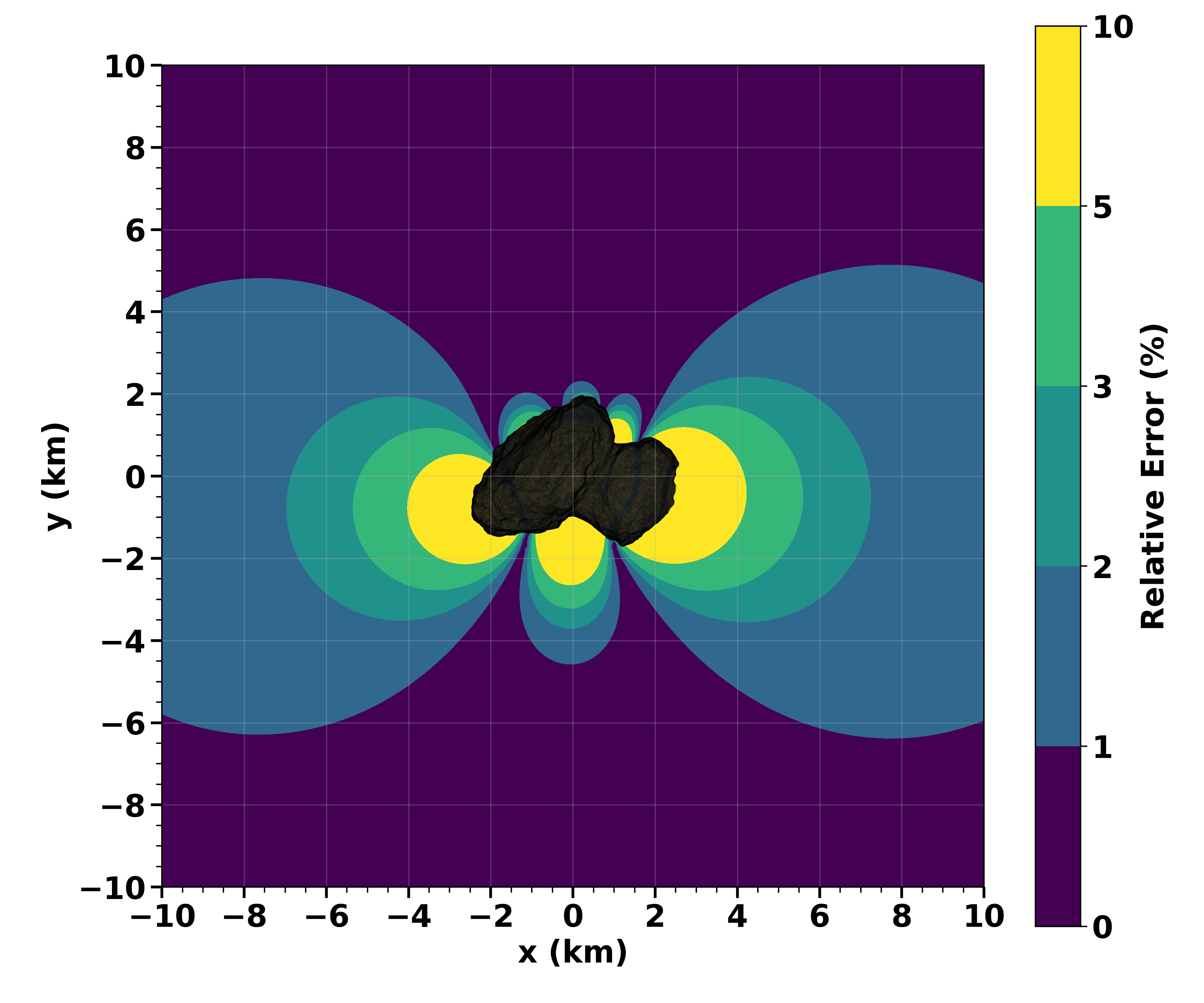}
    \caption{Gravitational potential relative error between Dipole Segment (DS) and polyhedron models. The colorbox indicates the relative error (\%) around comet 67P for points in the $xOy$ plane.}
    \label{errorrelative}
\end{figure}

\noindent Figure \ref{errorrelative} shows that, for approximated distances from the comet 67P between -4\,km and 4\,km, about the $x-$axis direction, the gravitational potential relative error between DS and polyhedron models is $>5\%$ (yellow). Also, for approximated distances from the comet 67P, between -3\,km and 2\,km, about the $y-$axis direction, the relative error is $>5\%$. On the other hand, for approximated distances from comet 67P, $<-4$\,km or $>4$\,km about the $x-$axis direction and $<-3$\,km or $>2$\,km about the $y-$axis direction, the relative error becomes $<5\%$, which suggests that the DS model can be used beyond this region for the study families of planar symmetric periodic orbits around comet 67P (sec. \ref{sec:4}).
%**********************************************************************
\section{Evolution of the Characteristic Curves in the Dipole Segment Model Approach}
\label{sec:4}
To investigate the evolution of the characteristic curves of comet 67P, based on the topological characterization of the orbits and families around the body \citep{Abadetal2023}, we use a simplified model of its shape, in which the gravitational potential is approximated by a Dipole Segment (DS).
The main goal of this section is to investigate the possibility of families of planar symmetric periodic orbits (SPO) around comet 67P, utilizing the DS model for this purpose. The main advantage of this simplification is that it overcomes the difficulty of finding periodic orbits, considering the irregular gravitational potential of the body, particularly for orbits far from the comet's surface.

\begin{figure}
\centering
\includegraphics[width=0.6\textwidth]{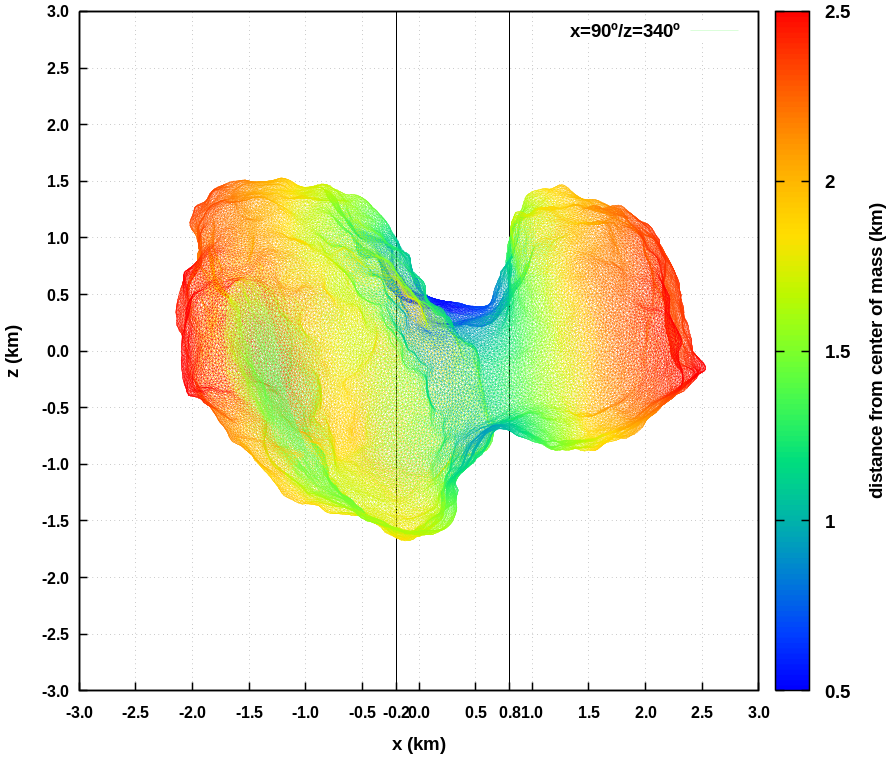}
\caption{3-D polyhedral shape model of the nucleus of comet 67P viewed from $xOz$ plane, rotating 90$^\circ$ around the $x-$axis and 340$^\circ$ around the $z-$axis. The color box indicates the surface distance from the center of mass (in km). The two black vertical lines indicate the considered polyhedral solid region for the comet's neck $\sim1$\,km (-0.2 to 0.8\, km).}
\label{fig:4:1}
\end{figure}

The DS models a massive segment connected by a mass at each end, similar to a ``bone'', which the shape of comet 67P fits approximately (see Fig. \ref{fig:4:1}).
In this model, the masses at the ends of the segment, considered to be spherical, are approximated as a point mass. We use the mass concentrations (mascons) constrained by the 3-D polyhedral surface to infer the masses of the lobes and the segment, as well as the segment length \citep{Zengetal2018, Elipeetal2021}. 

The parameters $k=\frac{GM}{\omega^2l^3}$, $\mu=\frac{M_2}{M_1+M_2}$ and $\mu_s=\frac{M_s}{M}$, being $M=M_1+M_2+M_s$, of the gravitational potential of the DS model (Eq. \ref{potDS}), in this case, were obtained from the 3-D polyhedral shape model filled with 151,906 points of mass concentration (mascons). 
\begin{equation}
U = -k \left[ \frac{(1 - \mu)(1 - \mu_s)}{r_1} + \frac{\mu(1 - \mu_s)}{r_2} + \mu_s \log\left( \frac{r_1 + r_2 + l}{r_1 + r_2 - l} \right) \right],
\label{potDS}
\end{equation}
where $r_1 = \sqrt{(x + \mu - \mu \mu_s + \mu_s/2)^2 + y^2 + z^2}
$ and 
$r_2 = \sqrt{(x + 1 + \mu - \mu \mu_s + \mu_s/2)^2 + y^2 + z^2}$.
We assumed from the comet 67P neck (Fig. \ref{fig:4:1}) that the segment has $\sim$1.0\,km (two black vertical lines) and mass $M_s$. So, we distributed 31,606 mascons in the neck polyhedral solid region between the two vertical lines in black (Fig. \ref{fig:4:1}), 76,440 mascons for the polyhedral big lobe region (the polyhedral solid region on the left side of the first black vertical line), and 43,860 mascons for the polyhedral small lobe region (the polyhedral solid region on the right side of the second black vertical line). After that, the parameters of the DS model are defined as shown in Table \ref{tab:3}.
\begin{table}
\caption{Parameters of comet 67P for DS model, being $M_1$, $M_2$, and $M_s$ the masses of the big lobe, small lobe, and segment, respectively. The size of the segment is defined by $l$, and $\omega$ is the angular speed.}
\label{tab:3}
\renewcommand{\arraystretch}{1.1}
\setlength{\tabcolsep}{5pt} 
\begin{tabular*}{\textwidth}{@{\extracolsep{\fill}}lccccccc}
\toprule
$M_1$ (kg) & $M_2$ (kg) & $M_s$ (kg) & $l$ (km) & $\omega$ (rad/s) & $\kappa$ & $\mu$ & $\mu_s$   \\
\midrule
$5.0\times10^{12}$ & $2.9\times10^{12}$ & $2.1\times10^{12}$ & 1.0 & $1.407\times10^{-4}$ & 33.6104 & 0.3645 & 0.2081 \\
\bottomrule
\end{tabular*}
\end{table}
The dimensionless equation of motion of a particle around the body is given by Eq. \ref{EqMotion}.
\begin{equation}
    \begin{aligned}
        \ddot{x}-2 \dot{y} & = x - \frac{\partial U}{\partial x}, \\
        \ddot{y}+2 \dot{x} & = y - \frac{\partial U}{\partial y}, \\
        \ddot{z} & = - \frac{\partial U}{\partial z}.
    \end{aligned}
    \label{EqMotion}
\end{equation}
The characteristic curves are the smooth, one-parameter, continuous curves that represent the families of SPO in the initial conditions map, in this case, calculated using the Grid Search Method \citep{Markellosetal1974}.

\begin{figure}
\centering
\includegraphics[width=1.0\textwidth]{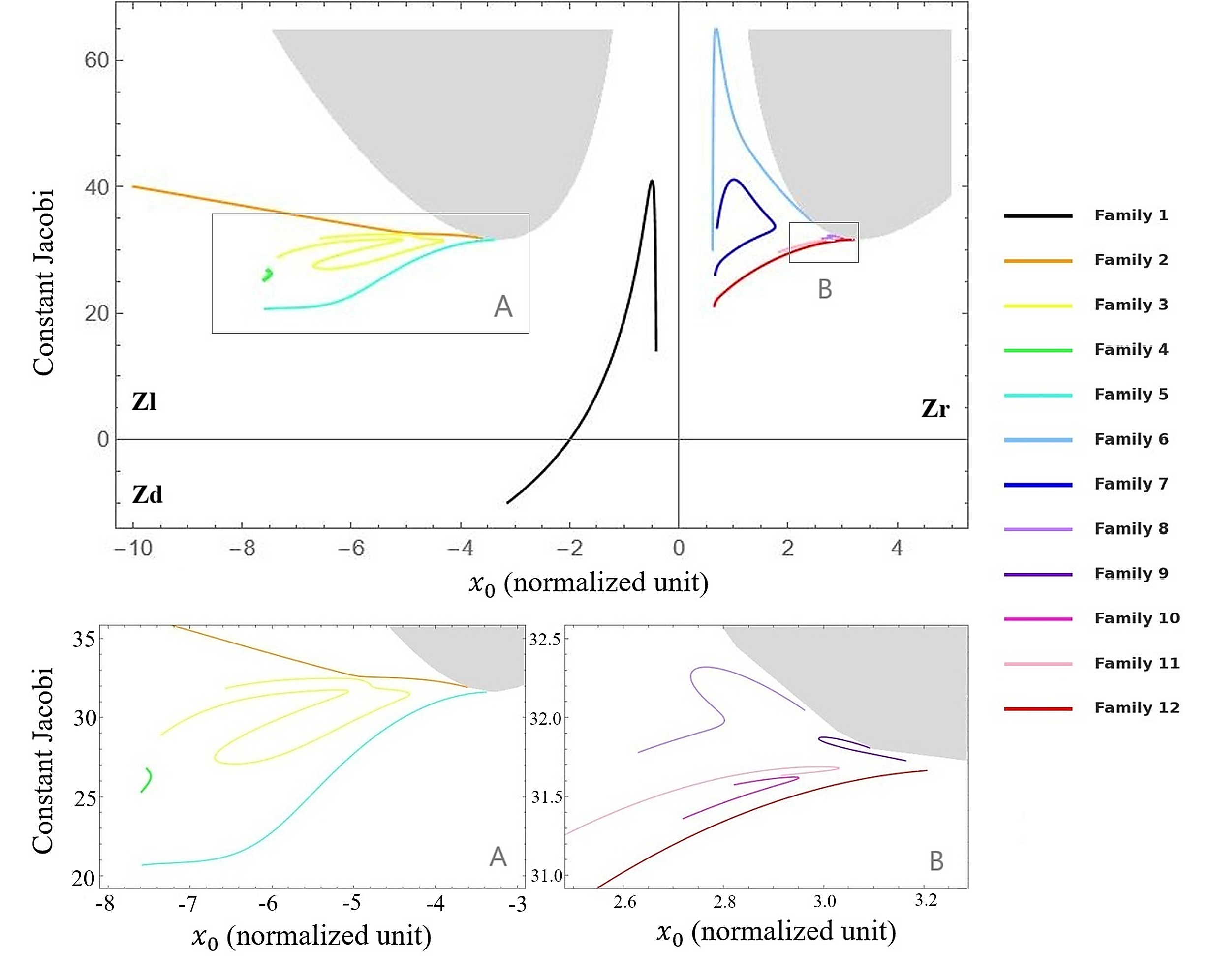}
\caption{Map of the characteristic curves of comet 67P. On the abscissa axis, we have the initial position $x_0$, and on the ordinate axis, the Jacobi constant $J$. $Z_l$, $Z_r$, and $Z_d$ name the quadrants where there are solutions. The colors of the lines represent each SPO family. The gray area represents the forbidden region, where for $y=0$, the condition $J > -(x^2 + y^2) + \frac{1}{2} U$ indicating a negative particle speed. The lower left figure is a closer view of the A rectangle in the $Z_l$ zone, and the lower right figure is a closer view of the B rectangle in the $Z_r$ zone.}
\label{MapaCI}
\end{figure}

For comet 67P, using $m=1$, $mtol=10^{-8}$, and $P=30$ periods, 12 SPO families were found, numbered from 1 to 12 on the initial conditions map, each represented by a color according to the legend (Fig. \ref{MapaCI}).
A point on the characteristic curve represents each orbit of the family.

Table \ref{Topologica} presents the topological classification of the characteristic curves that enables the identification of characteristics of orbital families without the need for direct visualization of the orbits \cite{Abadetal2023}. It shows the common features shared by all orbits within a family, as well as the variations occurring at the endpoints of each curve, based on the points encircled by the orbit, namely, the central body (denoted by $DS$) and the equilibrium points, whose locations and designations correspond to those in Fig. \ref{fig:8}, with $L_2$ and $L_4$ being the triangular points collectively represented by the single letter $T$.

This classification also accounts for the direction of motion: ``Direct'', denoted by $\mathcal{D(\quad)}$, for cases where $x_0 > x_{P/2}$, with $x_{P/2}$ defined as the $x$-position of the period $P/2$ where the curve intersects the $x-$axis perpendicularly, thereby characterizing a SPO; and ``Retrograde'', denoted by $\mathcal{R(\quad)}$, for cases where $x_0 < x_{P/2}$.
Consequently, the notation $[\quad]$ signifies that all orbits within the family exhibit similar characteristics; $[\rightarrow\;...]$ indicates that the orbits at one endpoint conform to the common family characterization while the other endpoint are different; and $[...\;\rightarrow\;...]$ implies that both endpoints differ from the common characterization. 

\begin{figure}
\centering
\subfloat[\textbf{Family 1}]{\includegraphics[width=0.45\textwidth]{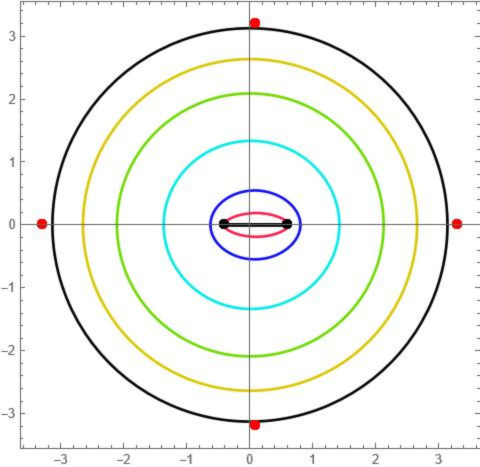}}
\hspace{1em}
\subfloat[\textbf{Family 2}]{\includegraphics[width=0.45\textwidth]{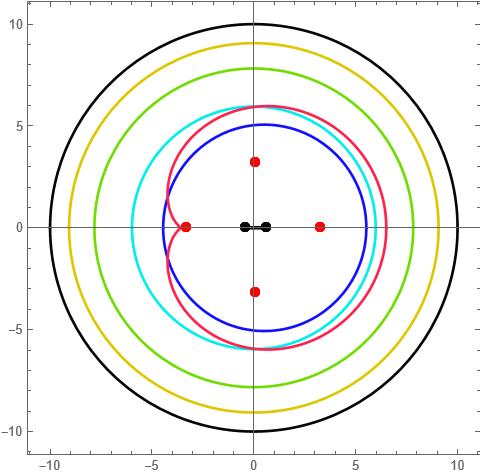}}\\
\subfloat[\textbf{Family 3}]{\includegraphics[width=0.45\textwidth]{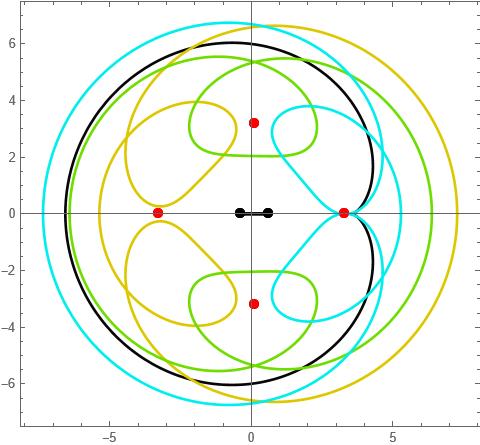}}
\hspace{1em}
\subfloat[\textbf{Family 4}]{\includegraphics[width=0.45\textwidth]{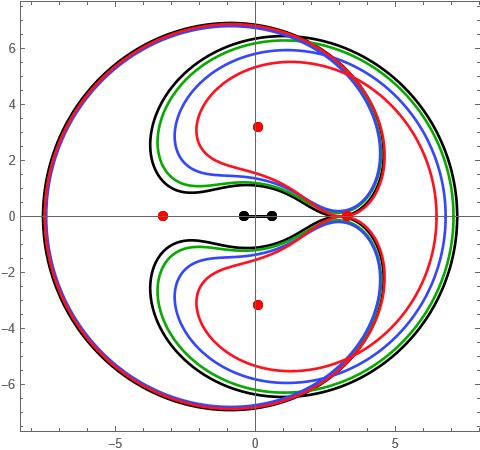}}
\caption{Evolution of the orbits of Families 1 to 4. Black dots with segments are the body (not to scale), and the red dots are the equilibrium points. The $x$ and $y$ axes represent the components of the particle's position around the body.}
\label{O1a4}
\end{figure}

\begin{figure}
\centering
\subfloat[\textbf{Family 5}]{\includegraphics[width=0.43\textwidth]{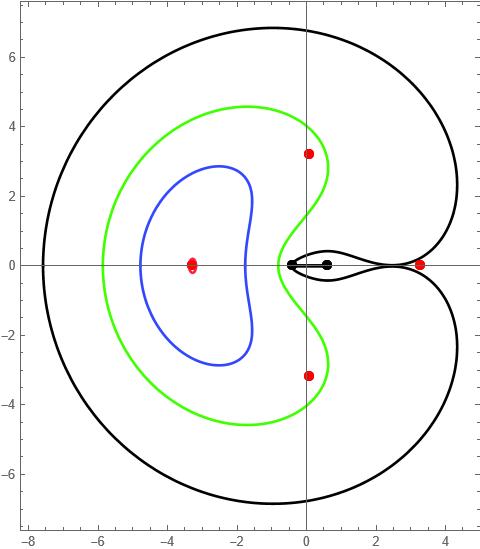}}
\hspace{1em}
\subfloat[\textbf{Family 6}]{\includegraphics[width=0.5\textwidth]{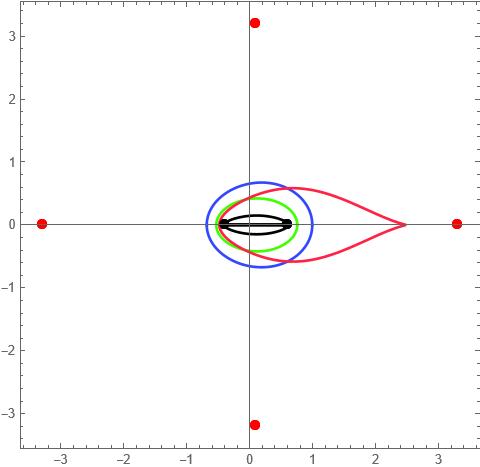}}\\
\subfloat[\textbf{Family 7}]{\includegraphics[width=0.43\textwidth]{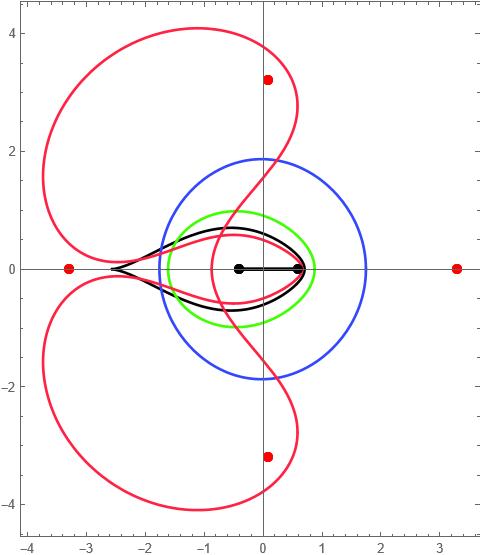}}
\hspace{1em}
\subfloat[\textbf{Family 8}]{\includegraphics[width=0.5\textwidth]{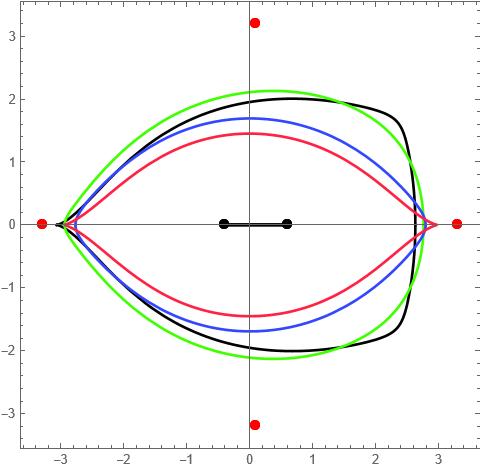}}
\caption{Evolution of the orbits of Families 5 to 8. Black dots with segments are the body (not to scale), and the red dots are the equilibrium points. The $x$ and $y$ axes represent the components of the particle's position around the body.}
\label{O5a8}
\end{figure}

\begin{figure}
\centering
\subfloat[\textbf{Family 9}]{\includegraphics[width=0.45\textwidth]{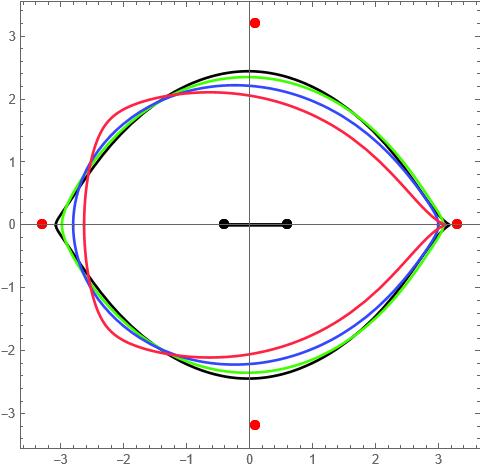}}
\hspace{1em}
\subfloat[\textbf{Family 10}]{\includegraphics[width=0.48\textwidth]{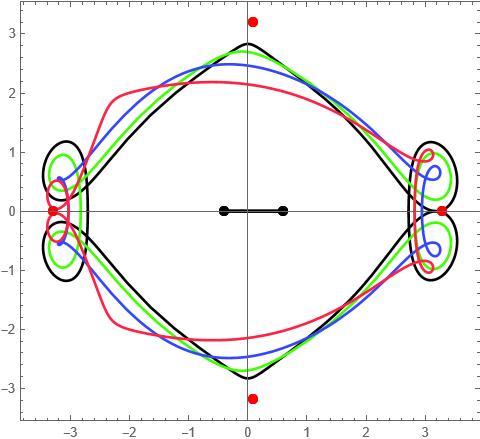}}\\
\subfloat[\textbf{Family 11}]{\includegraphics[width=0.52\textwidth]{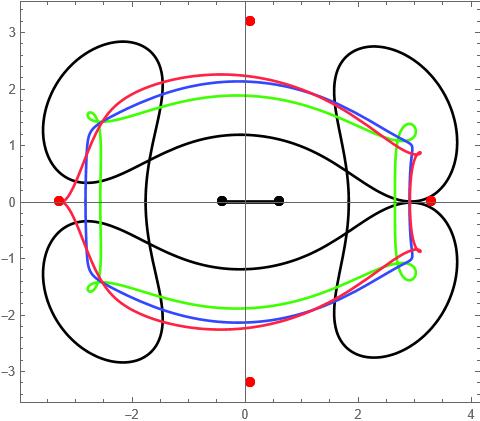}}
\hspace{1em}
\subfloat[\textbf{Family 12}]{\includegraphics[width=0.41\textwidth]{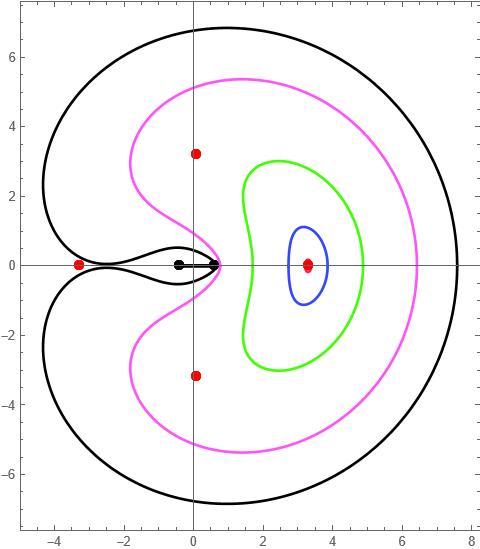}}
\caption{Evolution of the orbits of Families 9 to 12. Black dots with segments are the body (not to scale), and the red dots are the equilibrium points. The $x$ and $y$ axes represent the components of the particle's position around the body.}
\label{O9a12}
\end{figure}

Note, in Tab. \ref{Topologica}, that regions and corresponding families organize it, and the common part of the representation is highlighted in bold. Based on this classification and in the Figs. \ref{O1a4}, \ref{O5a8}, \ref{O9a12}, some analyses can be conducted:

\renewcommand{\arraystretch}{1.3} % Espaçamento vertical entre linhas
\begin{table}
\centering
\caption{Topological characterization of families of SPO for the comet 67P.}
\label{Topologica}
\begin{tabularx}{\textwidth}{l 
                                >{\centering\arraybackslash}X 
                                >{\centering\arraybackslash}X}
\toprule
\textbf{Region} & \textbf{Curves} & \textbf{Topological characterization} \\
\midrule
$Z_l$ and $Z_d$ & Family 1  & $\boldsymbol{R(DS)}\ [\rightarrow \{L_3 L_1 T\}]$ \\
$Z_l$           & Family 2  & $\boldsymbol{\mathcal{R}(L_3\ DS\ L_1\ T)}\ [\quad]$ \\
                & Family 3  & $\boldsymbol{\mathcal{R}(L_3\ DS\ L_1\ T)}\ [\quad]$ \\
                & Family 4  & $\boldsymbol{\mathcal{R}(L_3\ DS\ L_1\ T)}\ [\quad]$ \\
                & Family 5  & $\boldsymbol{\mathcal{R}(L_3)}\ [\rightarrow\ \{T\}]$ \\
$Z_r$           & Family 6  & $\boldsymbol{\mathcal{D}(DS)}\ [\quad]$ \\
                & Family 7  & $\boldsymbol{\mathcal{D}(DS)}\ [\rightarrow\ \{T\}]$ \\
                & Family 8  & $\boldsymbol{\mathcal{D}(DS)}\ [\quad]$ \\
                & Family 9  & $\boldsymbol{\mathcal{D}(DS)}\ [\quad]$ \\
                & Family 10 & $\boldsymbol{\mathcal{D}(DS)}\ [\quad]$ \\
                & Family 11 & $\boldsymbol{\mathcal{D}(DS)}\ [\quad]$ \\
                & Family 12 & $\boldsymbol{\mathcal{R}(L_1)}\ [\rightarrow\ \{T\}]$ \\
\bottomrule
\end{tabularx}
\end{table}
- Family 1 is the only one that covers two zones ($Z_l$ and $Z_d$), covering a wide range of the Jacobi constant. It represents elliptical curves around the body, which tend to be almost circular and possibly circular (Fig.~\ref{O1a4} (a)).

- From Family 2 to Family 5, the curves are in the $Z_l$ zone where $x_0<0$, fulfilling the requirement of $x_0<x_{P/2}$. In these cases, the orbits are retrograde. Some orbits of these families are represented in (Fig.~\ref{O1a4} (b), \ref{O1a4} (c), \ref{O1a4} (d) and \ref{O5a8} (a)).

- Families 2, 3, and 4  have the same topological characterization ($\boldsymbol{\mathcal{R}(L_3\ DS\ L_1\ T)}\ [\quad]$), they surround the body and all the equilibrium points, for the entire characteristic curve. The difference between them is in the looping that each one has (see Fig.~\ref{O1a4} (b), \ref{O1a4} (c), and \ref{O1a4} (d)).

- The orbits of the Families 6 to 12 are in the zone $Z_r$, where $x_0>0$, and direct rotation. Except Family 12 which meets $x_0<x_{P/2}$ and is retrograde. All of these cases have a similar common part, encircled the body (see Fig.~\ref{O5a8} (b), \ref{O5a8} (c), \ref{O5a8} (d), \ref{O9a12} (a), \ref{O9a12} (b), \ref{O9a12} (c) and \ref{O9a12}(d)).

- Families 6, 8, 9, 10 and 11 also have similar topological characterizations for the entire characteristic curve ($\boldsymbol{\mathcal{D}(DS)}\ [\quad]$) as in Fig. \ref{O5a8} (b), \ref{O5a8} (d), \ref{O9a12} (a), \ref{O9a12} (b) and \ref{O9a12} (c). Families 6, 8, and 9 are internal to all equilibrium points. The orbits of families 10 and 11 are close to the collinear points, but do not surround them.

- The orbits of the families 5 and 12 (\ref{O5a8} (a) and \ref{O9a12} (d)) initially encircling the collinear points $L_3$ and $L_1$ respectively, representing retrograde Lyapunov Orbits \textbf{($\mathcal{R}(L_3)$} and \textbf{$\mathcal{R}(L_1)$)}. At a certain point in the family's evolution, the orbits also surround the triangular points.

By comparing Figs. \ref{errorrelative} and \ref{O1a4}, we find that the DS model is suitable for investigating planar symmetric periodic orbits from Family 2 (with distances greater than 5 km) around comet 67P. This approach could be preferable to the polyhedron model, as it maintains a gravitational potential relative error of less than 5\%. This analysis could be helpful during specific orbital phases of space mission planning\footnote{\url{https://www.asteroidmission.org/asteroid-operations/}} \citep{ACCOMAZZO2015434}, as the use of the DS model reduces the computational effort required to calculate gravitational potential compared to the polyhedron method.

For instance, during the planning of the Rosetta mission, the distance from comet 67P was reduced to 9 km during the close approach phase. The orbital plane was shifted to the terminator plane for sufficient confidence in the dynamical stability of the comet's environment. This strategy was essential to favorable observation conditions with the necessity to minimize the spacecraft's exposure to the predominantly radial gas flow emitted by the comet \citep{ACCOMAZZO2015434}. 

As shown in Fig. \ref{errorrelative}, we could employ the DS model to analyze the orbital dynamics around comet 67P in the close approach phase at the terminator plane, since it yields a gravitational relative error of less than 2\% compared with the polyhedron one. Furthermore, for other orbital phases, such as comet characterization, the relative errors are even smaller \citep{ACCOMAZZO2015434}.
%
%%%% Stability Index %%%%%%
\subsection{Stability of the Symmetric Periodic Orbits}
\label{sec:stability}
An orbit is linearly stable when applied to small perturbations around a periodic solution. From the linearization of the equations of motion (Eq.~\eqref{EqMotion}), the equilibrium of forces, predictability, and periodicity of the orbit are maintained. The linear stability is global and considers all directions in phase space.

The stability is determined from the eigenvalues of the monodromy matrix $M = \Phi(t_0+T,t_0)$ associated with the orbit, since $\dot \Phi = A(t)\Phi(t,t_0)$ is the integrated state transition matrix, $t_0$ is the initial time and $T$ is the complete period of the orbit; $A(t) \in \mathbb{R}^{6 \times 6}$ is the Jacobian matrix of the system and the state transition matrix at the initial time equal to the identity matrix \citep{parker2014low}. 

Because it is a Hamiltonian system, the matrix $M$ has six eigenvalues that, due to the symplectic structure of the system, appear in reciprocal pairs and, as a consequence of the periodicity of the orbit and energy conservation, two of them are equal to one, therefore $\lambda_1, \lambda_2=\frac{1}{\lambda_1}, \lambda_3, \lambda_4=\frac{1}{\lambda_3}, \lambda_5=1, \lambda_6=1$. 

\vspace{1em}
\noindent Horizontal Stability - Since horizontal stability refers to the response of the orbit to small perturbations within the orbital plane, the Jacobian matrix for this case reduces to the subspace $(x,\dot x,y,\dot y)$ as in Eq.~\eqref{MatA}. $V_{xx}, V_{xy}, V_{yx}, V_{yy}$ the second derivative of the effective potential (Eq.~\eqref{eq:1}).
\begin{align}
A_{h}(t)= \left[\begin{array}{cccc}
0 & 0 & 1 & 0 \\
0 & 0 & 0 & 1 \\
V_{xx} & V_{xy} & 0 & 2 \\
V_{xy} & V_{yy} & -2 & 0
\end{array}\right]. \label{MatA}
\end{align}
Therefore, the characteristic polynomial is
\begin{align}
P(\lambda_h)=\lambda_h^4-(S_h-2)\lambda_h^3+(2+2S_h)\lambda_h^2-(2+S_h)\lambda_h+1, \label{PolCMM}
\end{align}
where $S_h=\mid \lambda_{h1}+\frac{1}{\lambda_{h1}}\mid$ is the horizontal stability index, a single metric for assessing the stability of the system \citep{parker2014low}. If $S_h > 2$, the orbit is unstable; $S_h = 2$, the orbit is neutrally stable; and $S_h < 2$, the orbit is stable.

\vspace{1em}
\noindent Vertical Stability - A planar periodic orbit in a three-dimensional system is vertically stable if, when a small out-of-plane perturbation is applied, the motion remains bounded—that is, the body oscillates around the plane and does not drift away indefinitely. 

Mathematically, the planar periodic orbit is vertically stable if the eigenvalues of the monodromy matrix $\Phi_{v}(T)$, $\lambda_{v1}$ and $\lambda_{v2}$, which are associated with the out-of-plane behavior, lie on the unit circle of the complex plane (i.e., $\mid \lambda_{v1}\mid = \mid \lambda_{v2}\mid = 1$ ). Otherwise, it is vertically unstable. Since $S_v=\mid \lambda_{v1}+\lambda_{v2}\mid = \mid \lambda_{v1}+\frac{1}{\lambda_{v1}}\mid$, when the limit of vertical stability occurs we have $S_v=2$.

\begin{figure}[h!]
\centering
    \subfloat[]{\includegraphics[width=0.4\textwidth]{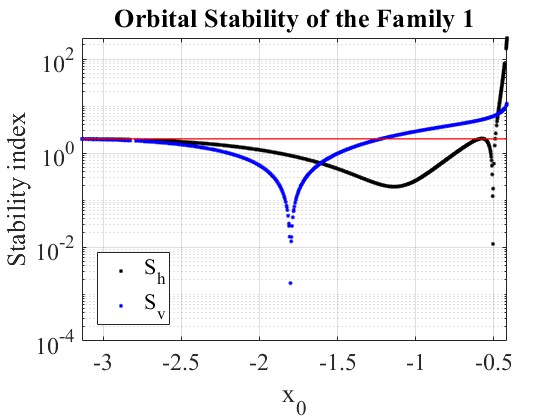}}
   \hspace{1em} % Adiciona espaço entre as figuras
   \subfloat[]{\includegraphics[width=0.4\textwidth]{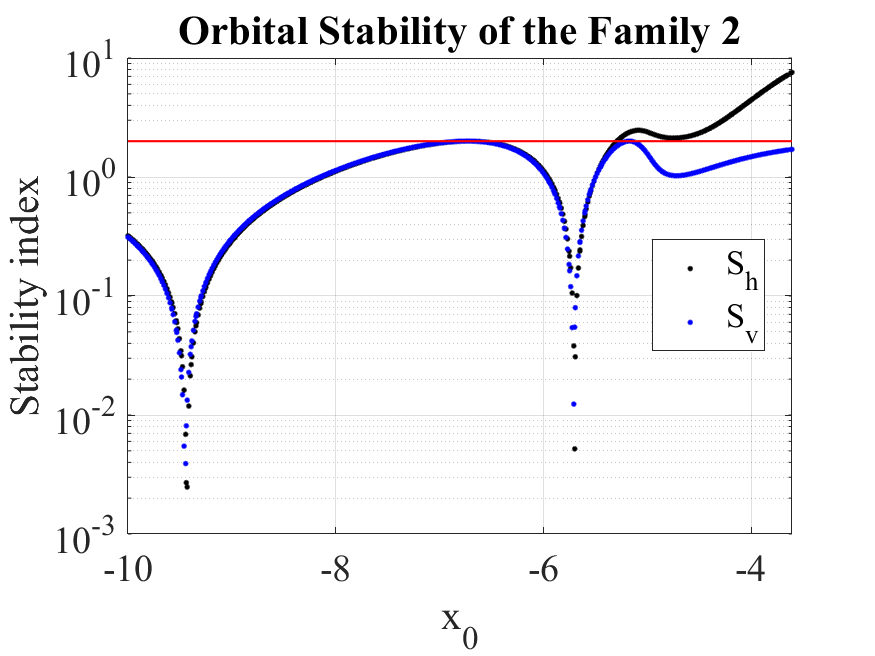}}
   \hspace{1em} % Adiciona espaço entre as figuras
  \subfloat[]{\includegraphics[width=0.4\textwidth]{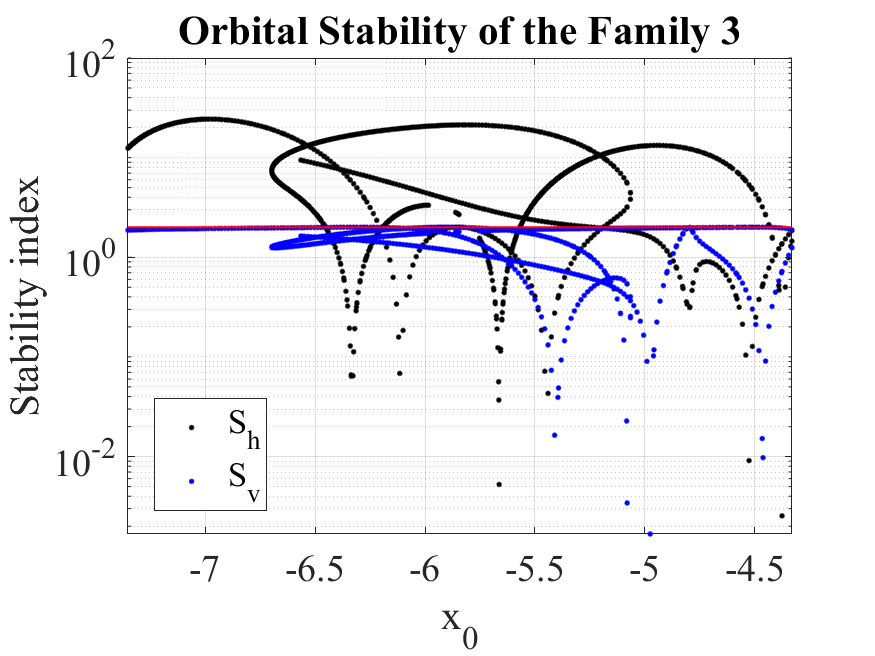}}
     \hspace{1em} % Adiciona espaço entre as figuras
   \subfloat[]{\includegraphics[width=0.4\textwidth]{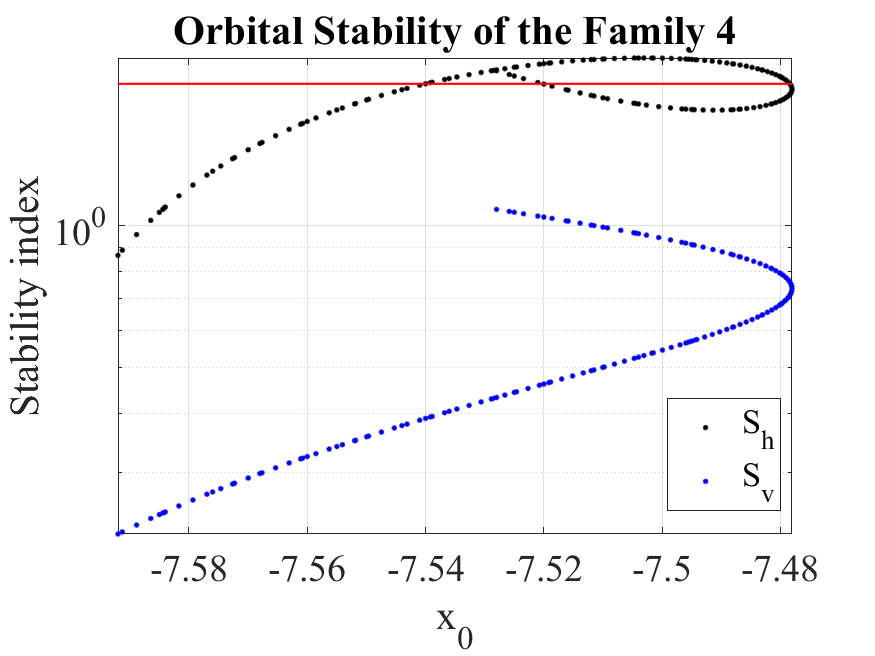}}
       \hspace{1em} % Adiciona espaço entre as figuras
   \subfloat[]{\includegraphics[width=0.4\textwidth]{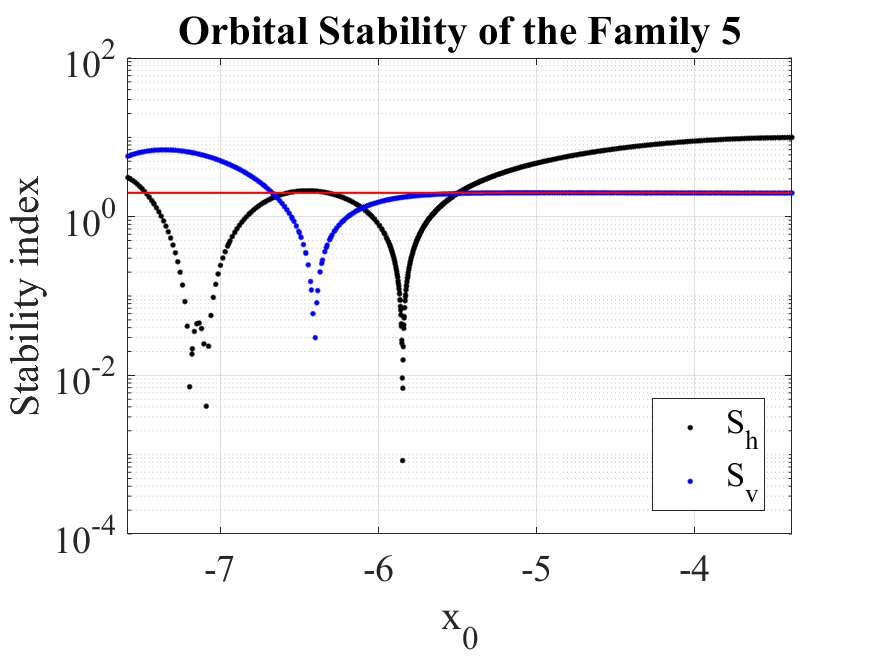}}
       \hspace{1em} % Adiciona espaço entre as figuras
    \subfloat[]{\includegraphics[width=0.4\textwidth]{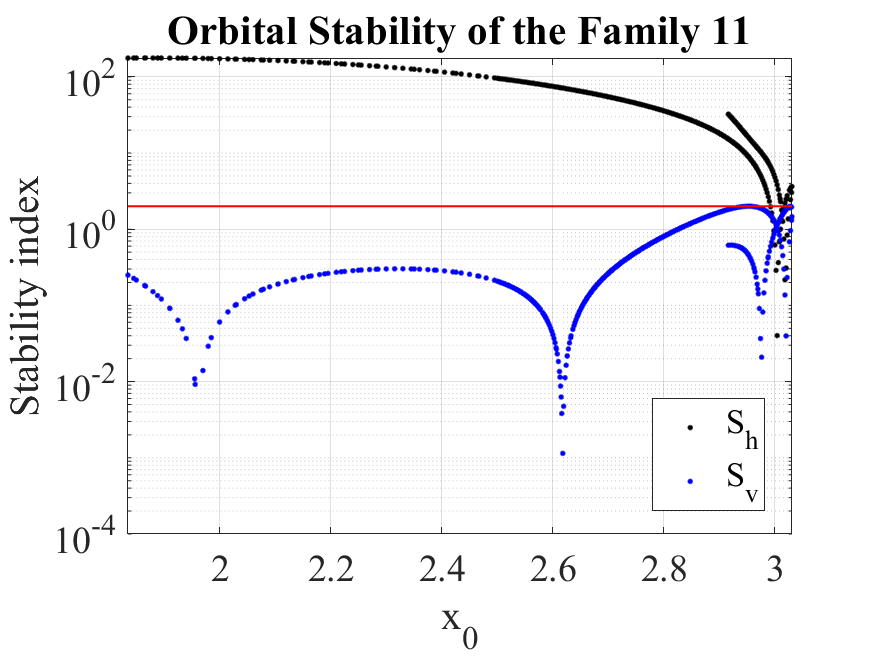}}
       \hspace{1em} % Adiciona espaço entre as figuras
   \subfloat[]{\includegraphics[width=0.4\textwidth]{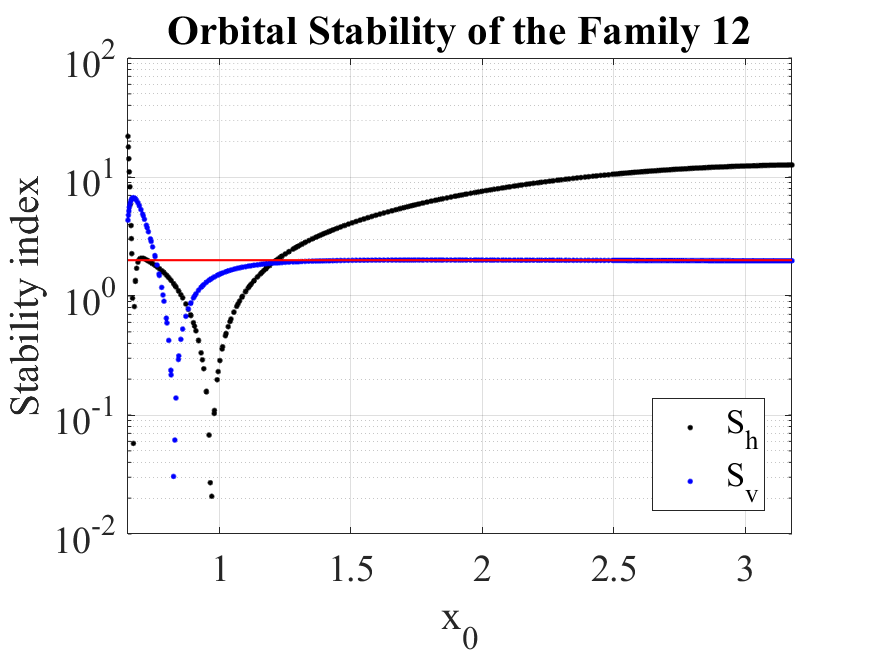}}
   \hspace{1em} % Adiciona espaço entre as figuras
      \caption{Stability index referring to families: (a) 1, (b) 2, (c) 3, (d) 4, (e) 5, (f) 11, (g) 12. The red line continues at a stability index equal $2.0$.}
    \label{fig:stability}
\end{figure}

The Jacobian matrix reduce to the subspace $(z,\dot z)$, as in Eq.~\eqref{MatB}, where $V_{zz}$ the second derivative of the effective potential (Eq.~\eqref{eq:1}), calculated for $z=0$ since the orbit is planar. 
\begin{align}
A_{v}(t)= \left[\begin{array}{cc}
0 & 1  \\
V_{zz} & 0  \\
\end{array}\right]. \label{MatB}
\end{align}
Figure \ref{fig:stability} shows the horizontal (black points) and vertical (blue points) stability index for some SPO families presented in the initial conditions map (Fig. \ref{MapaCI}). The solid red line is at a stability index equal $2.0$.

Note in Fig. \ref{fig:stability}, that in the families we analyzed there is a change from horizontal stability to instability, passing through index $2.0$, this means that $\lambda_{h1}$ and $\lambda_{h2}$ are $\pm1$, which implies critical points where bifurcations (exchange of horizontal stability) can occur.

Regarding vertical stability, which defines the stability outside the plane of the planar orbit, note that in the Families $1$, $5$ and $12$ there are stable and unstable orbits; in Families $2$, $4$ and $11$ only stability; and in Family $3$ the maximum index is $S_v=2.000398$ (Fig. \ref{fig:stability}).

When $S_v=2$, the planar family of periodic orbits bifurcates into a 3D family of periodic orbits, maintaining the multiplicity (for $\lambda_{v1}=1$) or doubling it ($\lambda_{v1}=-1$), depending on the sign of $S_v$.

If $0<S_v<2$ and $S_v/2 = |cos(2\pi m/n)$ with $m,n \in \mathbb{Z}$ and $m<n$, then the SPO family can bifurcate into a 3D family of periodic orbits with multiplicity $nM$.

 \vspace{1em}
\noindent Validation - Since the system reduces to a 2D Hamiltonian problem for the vertical stability problem, some information can be analyzed to validate the results:

- The determinant of the monodromy matrix must be equal to $1$;

- If the eigenvalues are complex conjugates of magnitude $1$, the system is stable; if they are real reciprocals, the system is unstable.

- For systems with a monodromy matrix $\mathbb{R}^{2 \times 2}$ and a determinant equal to $1$, $S_v =\mid \lambda_{v1}+\lambda_{v2}\mid = \mid \lambda_{v1}+\frac{1}{\lambda_{v1}}\mid= \mid\operatorname{tr}{(\Phi_v(T))\mid}$, since $\operatorname{tr}$ is the trace of the monodromy matrix .

- A small vertical perturbation of $z_0 = 0.001$ was used to evaluate vertical stability.

- The vertical stability was calculated using the complete trajectory $x(t), y(t)$ of the periodic orbit.

\vspace{5em}

\section{Conclusion}
\label{sec:5}

This work presented a comprehensive study of the surface and orbital dynamics of comet 67P/Churyumov-Gerasimenko using a detailed 3-D polyhedral shape model. We treated the comet's nucleus as a surface similar to that of an asteroid, since no drag force was considered in our analysis. Our objective was to investigate the dynamic features of the surface, particularly the slopes, while incorporating Third-Body (TB) and Solar Radiation Pressure (SRP) perturbations. Additionally, we examined the orbital dynamics around the comet using the DS model, with a focus on planar symmetric periodic orbits.

We observed that the comet's surface presents 98.5\% of tilts lower than 100$^{\circ}$. The analysis of its geopotential revealed that the gravitational potential predominates over the centrifugal potential, due to the comet's slow rotation.

The big lobe exhibits maximum surface acceleration values. In contrast, the Hapi (comet's neck) region exhibits intermediate surface acceleration values, indicating a more substantial influence of gravitational potential in the big lobe than in the Hapi region, closer to the comet's center of mass. We also found that the surface acceleration of comet 67P has its maximum value, $\sim$2.2$\times$10$^{-7}$ km\,s$^{-2}$, in its big lobe.

Comparing our results with previous works, we found that the slopes are particularly noteworthy in the intervals from 20$^{\circ}$ to 40$^{\circ}$ and 60$^{\circ}$ to 90$^{\circ}$.

Regarding the escape speed, it is maximum in the Hapi region, a common feature with other slowly rotating bodies. Additionally, the geopotential minimum regions are the areas with the maximum escape speed.

Additionally, we examined the impact of Third-Body gravitational effects and Solar Radiation Pressure on slope behavior. Our findings indicate that Third-Body perturbations have minimal influence on the overall behavior of the slopes. Conversely, Solar Radiation Pressure also appears to have a negligible effect on particles on the surface of comet 67P, specifically for sizes greater than approximately \(10^{-3}\) cm at apocenter and greater than approximately \(10^{-1}\) cm at pericenter.

We identified five equilibrium points considering the current density and rotational period of comet 67P. Different from previous works, we found that the outer equilibrium point E$_2$ and the inner equilibrium point E$_5$ are linearly stable. The other equilibrium points are unstable. In addition, the Roche lobe (teardrop-shaped region, Fig. \ref{fig:8}) intersects at the equilibrium point E$_1$.

To analyze the orbital evolution around comet 67P, we simplified the irregular shape of the body using the DS model. A topological classification of the orbit families was performed, revealing the dynamics around comet 67P based on the characteristics of the families, without requiring direct visualization of the orbits. We found 12 families of planar symmetric periodic orbits around the body. We also investigated the horizontal and vertical stabilities of these families.

Finally, comparing the DS and polyhedron models, our results suggest that the DS model could be used for the investigation of families of planar symmetric periodic orbits that have distances larger than $\sim5$\,km from comet 67P. For this case, the gravitational potential relative error compared with the polyhedron method is $<5\%$.

%Finally, this work presented the study of the surface and orbital dynamics of comet 67P, both important dynamic aspects to provide an initial theoretical basis to aid the identification of more economical trajectories for parking a spacecraft.

%\textbf{Finally, this work presented the study of the surface and orbital dynamics of comet 67P, both important dynamic aspects for the investigation of similar small bodies for future space missions.}

\bmhead{Acknowledgements} The authors thank the São Paulo State University (UNESP), School of Engineering and Sciences, Campus Guaratinguetá (grant 06/2023-PROPe - IEPe-RC), the Unesp Institutional Scientific Initiation Scholarship Program - PIBIC (grant PROPe Unesp 09/2023), the Center for Scientific Computing (NCC/GridUNESP) of the São Paulo State University (UNESP), the Center for Mathematical Sciences Applied to Industry (CeMEAI), funded by FAPESP (grant 2013/07375-0), and FAPESP, for financial support (proc. 2023/11781-5).

%Acknowledgements are not compulsory. Where included they should be brief. Grant or contribution numbers may be acknowledged.

%Please refer to Journal-level guidance for any specific requirements.

%\section*{Declarations}

%Some journals require declarations to be submitted in a standardised format. Please check the Instructions for Authors of the journal to which you are submitting to see if you need to complete this section. If yes, your manuscript must contain the following sections under the heading `Declarations':

%\item Ethics approval and consent to participate
%\item Consent for publication
\bmhead{Author Contribution Statement} LB, AA, AF, CG, and LC equally contributed to this work. LB, AA, AF, CG, and LC performed numerical simulations, figures, and wrote the manuscript. AA and AF revised the manuscript.
\bmhead{Data availability} Data will be made available on a reasonable request to the corresponding author.
\bmhead{Code availability} The numerical codes used in this work can be found at GitHub: \url{https://github.com/a-amarante}
\bmhead{Conflict of interest} The authors declare no conflict of interest.

\bibliography{sn-bibliography}% common bib file


\end{document}